\documentclass[submission, PhysCodeb]{SciPost}

\pdfoutput=1

\usepackage[utf8]{inputenc} 
\usepackage[T1]{fontenc} 	
\usepackage[english]{babel} 

\usepackage[bitstream-charter]{mathdesign}
\usepackage{geometry} 		
\usepackage{amsmath} 		
\usepackage{mathtools} 		
\usepackage{float} 			
\usepackage{graphicx} 		
\usepackage{tabularx} 		
\usepackage{booktabs} 		
\usepackage{color, xcolor} 	
\usepackage{pdfpages} 		
\usepackage{extarrows} 		
\usepackage{multirow} 		
\usepackage{multicol} 		
\usepackage{caption} 		
\usepackage{comment}
\usepackage{subcaption} 	
\usepackage{enumitem} 		
\usepackage{xspace} 		
\usepackage{stackrel} 		
\usepackage{tikz} 			
\usetikzlibrary{calc}
\usepackage{braket} 		
\usepackage{bm} 			
\usepackage{tensor} 		
\usepackage{slashed} 		
\usepackage{siunitx} 
\usepackage{lastpage} 		
\usepackage{cite} 			
\usepackage[normalem]{ulem} 
\usepackage{fontawesome} 	
\usepackage{tocloft} 		
\usepackage{titlesec} 		
\usepackage{doi} 			
\usepackage[most]{tcolorbox} 					
\usepackage[capitalize]{cleveref} 	
\usepackage[nottoc, notlot, notlof]{tocbibind} 	
\usepackage[ruled, vlined]{algorithm2e} 		
\usepackage{makecell}
\usepackage{pifont}

\hypersetup{
	pdftitle={Agentic Re-Casting using Agentic Re-Simulations},
	pdfauthor={Sascha Diefenbacher, Tilman Plehn, Daniel Schiller, and Nikita Schmal},
    linktoc=section,
	breaklinks=true,			
	colorlinks=true, 			
	linkcolor={red!50!black}, 	
	citecolor={blue!50!black}, 	
	urlcolor={blue!80!black} 	
} 

\DeclareSymbolFont{usualmathcal}{OMS}{cmsy}{m}{n}
\DeclareSymbolFontAlphabet{\mathcal}{usualmathcal}

\SetArgSty{textnormal}
\SetKwComment{Comment}{{\small\#}~}{}
\SetCommentSty{mycommfont}

\setitemize{itemsep=2pt, parsep=0pt} 				
\setenumerate{itemsep=2pt, parsep=0pt} 				
\crefname{section}{Sec.}{Secs.}
\crefname{equation}{Eq.}{Eqs.}
\crefname{figure}{Fig.}{Figs.}
\crefname{table}{Tab.}{Tabs.}
\Crefname{section}{Section}{Sections}
\Crefname{equation}{Equation}{Equations}
\Crefname{figure}{Figure}{Figures}
\Crefname{table}{Table}{Tables}

\usepackage[most]{tcolorbox}
\tcbuselibrary{minted,skins,breakable}

\usepackage{dirtree}

\newcommand{\eqc}{\;\text{,}} 		

\definecolor{sci_light_blue}{HTML}{4869b4}
\definecolor{sci_light_tes}{HTML}{a08aaa}
\definecolor{sci_light_gre}{HTML}{3E5F58}
\definecolor{red_cb}{HTML}{e41a1c}
\definecolor{blue_cb}{HTML}{377eb8}
\definecolor{green_cb}{HTML}{4daf4a}
\definecolor{purple_cb}{HTML}{984ea3}
\definecolor{orange_cb}{HTML}{ff7f00}

\newcommand{\mav}[1]{\textsc{MadAgents.v{#1}}\xspace}
\newcommand{\ma}{\textsc{MadAgents}\xspace}

\newcommand{\madgraph}{\textsc{MadGraph5\_aMC@NLO}\xspace}
\newcommand{\mg}{\textsc{MadGraph}\xspace}
\newcommand{\sfa}{\textsc{SFitter} agents\xspace}

\newcommand{\sfitter}{\textsc{SFitter}\xspace}

\newcommand{\mwith}{\text{with}}

\newcommand{\qqquad}{\qquad\quad}

\newcommand{\pois}{\text{Poiss}}

\newcommand{\ope}{\mathcal{O}} 			    

\newcommand{\ttbar}{t\bar{t}} 			

\newcommand{\arXiv}[2][]{%
	\ifthenelse{\equal{#1}{}}%
	{\href{http://arxiv.org/abs/#2}{arXiv:#2}}%
	{\href{http://arxiv.org/abs/#2}{arXiv:#2~[#1]}}}

\newcommand{\gev}{\text{GeV}}
\newcommand{\tev}{\text{TeV}}

\newcommand{\checksquare}{\makebox[0pt][l]{$\square$}\raisebox{.15ex}{\hspace{0.1em}$\checkmark$}}

\def\slashchar#1{\setbox0=\hbox{$#1$}           
   \dimen0=\wd0                                 
   \setbox1=\hbox{/} \dimen1=\wd1               
   \ifdim\dimen0>\dimen1                        
      \rlap{\hbox to \dimen0{\hfil/\hfil}}      
      #1                                        
   \else                                        
      \rlap{\hbox to \dimen1{\hfil$#1$\hfil}}   
      /                                         
   \fi}

\newcommand{\tikznode}[2]{%
\ifmmode%
\tikz[remember picture,baseline=(#1.base),inner sep=0pt] \node (#1) {$#2$};%
\else
\tikz[remember picture,baseline=(#1.base),inner sep=0pt] \node (#1) {#2};%
\fi}

\def\mathswitchr#1{\relax\ifmmode{\text{#1}}\else$\text{#1}$\xspace\fi}
\def\mathswitch#1{\relax\ifmmode#1\else$#1$\xspace\fi}

\makeatletter
\newcommand{\indentafterblock}{%
  \futurelet\pb@next\pb@after
}
\newcommand{\pb@after}{%
  \ifx\pb@next\par
  \else
    \ifhmode\par\fi
    \noindent\ignorespaces
  \fi
}
\makeatother

\newtcblisting{bashblock}{
  listing engine=minted,
  minted language=bash,
  breakable,
  enhanced,
  colback=black!2,
  colframe=black!20,
  arc=2pt,
  boxrule=0.45pt,
  left=4pt,right=4pt,top=2pt,bottom=2pt,
  boxsep=1pt,
  listing only,
  before skip=4pt, after skip=4pt,
  after=\indentafterblock,
  minted options={
    breaklines,
    breaksymbolleft=,
    fontsize=\footnotesize,
    autogobble,
    tabsize=2
  }
}

\tcbuselibrary{skins,breakable,listings}

\tcbset{
  promptblock/base/.style={
    breakable,
    enhanced,
    arc=2pt,
    boxrule=0.45pt,
    left=4pt,right=12pt,top=4pt,bottom=3pt,
    boxsep=1pt,
    title={#1},
    title filled,
    coltitle=black,
    fontupper=\footnotesize,
    fonttitle=\ttfamily\small\bfseries,
    toptitle=1.5pt,bottomtitle=1.5pt,
    lefttitle=4pt,righttitle=4pt,
    before skip=4pt, after skip=4pt,
    after=\indentafterblock
  }
}

\tcbset{
  promptblock/plan/.style={
    colback=sci_light_tes!5,
    colframe=sci_light_tes!50,
    colbacktitle=sci_light_tes!50
  },
  promptblock/user/.style={
    colback=sci_light_gre!5,
    colframe=sci_light_gre!50,
    colbacktitle=sci_light_gre!50 
},
  promptblock/final/.style={
    colback=sci_light_blue!5,
    colframe=sci_light_blue!50,
    colbacktitle=sci_light_blue!50
  }
}

\newtcolorbox{userblock}[1][]{
  promptblock/base={#1},
  promptblock/user
}

\newtcolorbox{planblock}[1][]{
  promptblock/base={#1},
  promptblock/plan
}

\newtcolorbox{finalblock}[1][]{
  promptblock/base={#1},
  promptblock/final
}

\newtcolorbox{promptblock}[1][]{
  breakable,
  enhanced,
  colback=sci_light_blue!5,
  colframe=sci_light_blue!50,
  arc=2pt,
  boxrule=0.45pt,
  left=4pt,right=12pt,top=4pt,bottom=3pt,
  boxsep=1pt,
  title={#1},
  title filled,
  colbacktitle=sci_light_blue!50,
  coltitle=black,
  fontupper=\footnotesize,
  fonttitle=\ttfamily\small\bfseries,
  toptitle=1.5pt,bottomtitle=1.5pt,
  lefttitle=4pt,righttitle=4pt,
  before skip=4pt, after skip=4pt,
  after=\indentafterblock
}

\newtcblisting{promptblocklit}[1][]{
  breakable,
  enhanced,
  colback=black!3,
  colframe=black!35,
  arc=2pt,
  boxrule=0.45pt,
  left=2pt,right=2pt,top=2pt,bottom=2pt,
  boxsep=0.5pt,
  title={#1},
  title filled,
  colbacktitle=black!20,
  coltitle=black,
  fonttitle=\ttfamily\footnotesize\bfseries,
  toptitle=1pt,bottomtitle=1pt,
  lefttitle=2pt,righttitle=2pt,
  before skip=3pt, after skip=3pt,
  listing only,
  listing engine=listings,
  listing options={
    basicstyle=\ttfamily\tiny\linespread{0.95}\selectfont,
    columns=fullflexible,
    keepspaces=true,
    showspaces=false,
    showstringspaces=false,
    breaklines=true,
    breakatwhitespace=false,
    upquote=true,
    aboveskip=0pt,
    belowskip=0pt,
    literate={—}{{---}}1 {“}{{``}}1 {”}{{''}}1 {→}{{->}}2 {–}{{--}}1 {‘}{{`}}1 {’}{{'}}1 {…}{{...}}3 {√}{{sqrt}}4
  },
  after=\indentafterblock
}

\newtcblisting{promptblocklitlarge}[1][]{
  breakable,
  enhanced,
  colback=black!3,
  colframe=black!35,
  arc=2pt,
  boxrule=0.45pt,
  left=2pt,right=2pt,top=2pt,bottom=2pt,
  boxsep=0.5pt,
  title={#1},
  title filled,
  colbacktitle=black!20,
  coltitle=black,
  fonttitle=\ttfamily\footnotesize\bfseries,
  toptitle=1pt,bottomtitle=1pt,
  lefttitle=2pt,righttitle=2pt,
  before skip=3pt, after skip=3pt,
  listing only,
  listing engine=listings,
  listing options={
    basicstyle=\ttfamily\footnotesize\linespread{0.95}\selectfont,
    columns=fullflexible,
    keepspaces=true,
    showspaces=false,
    showstringspaces=false,
    breaklines=true,
    breakatwhitespace=false,
    upquote=true,
    aboveskip=0pt,
    belowskip=0pt,
    literate={—}{{---}}1 {“}{{``}}1 {”}{{''}}1 {→}{{->}}2 {–}{{--}}1 {‘}{{`}}1 {’}{{'}}1 {…}{{...}}3 {√}{{sqrt}}4
  },
  after=\indentafterblock
}

\usepackage[fencedCode,pipeTables,texMathDollars,tightLists,frozenCache,cacheDir=md-cache]{markdown}

\AddToHook{begindocument/end}{%
  \markdownSetup{
    renderers = {
      softLineBreak = {\unskip\space},
      hardLineBreak = {\unskip\space},
    },
  }%
}

\usepackage{fvextra}
\usepackage{xurl}
\usepackage{xstring}

\usetikzlibrary{shadows.blur}

\definecolor{mdAccent}{HTML}{2563EB} 
\definecolor{mdBg}{HTML}{F8FAFC}     
\definecolor{mdFrame}{HTML}{CBD5E1}  
\definecolor{mdTitleBg}{HTML}{EFF6FF}

\tcbset{
  mdCard/.style={
    enhanced,
    breakable,
    colback=mdBg,
    colframe=mdAccent!35!mdFrame,
    arc=2.2mm,
    boxrule=0.45pt,
    left=1.8mm,right=1.8mm,top=1.3mm,bottom=1.3mm,
    boxsep=0.9mm,
    colbacktitle=mdTitleBg,
    coltitle=black,
    fonttitle=\ttfamily\small\bfseries,
    boxed title style={sharp corners, boxrule=0pt},
    toptitle=0.8mm,bottomtitle=0.8mm,
    before skip=4pt,
    after skip=4pt
  }
}

\newtcolorbox{mdcodeblock}[1][]{%
  enhanced,
  breakable,
  colback=mdBg,
  colframe=mdAccent!30!mdFrame,
  arc=2.0mm,
  boxrule=0.4pt,
  left=4.5pt,right=4.5pt,top=2.2pt,bottom=2.2pt,
  boxsep=1pt,
  #1
}

\makeatletter

\def\markdownRendererCodeSpan#1{%
  {\ttfamily\footnotesize
    \begingroup
      \def\markdownRendererUnderscore##1{\string_}%
      \def\markdownRendererBackslash##1{\string\\}%
      \def\markdownRendererLeftBrace##1{\string\{}%
      \def\markdownRendererRightBrace##1{\string\}}%
      \def\markdownRendererDollarSign##1{\string$}%
      \def\markdownRendererTilde##1{\string~}%
      \def\markdownRendererCircumflex##1{\string^}%
      \edef\md@codespan{#1}%
      \expandafter\path\expandafter{\md@codespan}%
    \endgroup
  }%
}

\makeatother

\newcommand{\catrule}[1]{%
  \multicolumn{2}{@{}l@{}}{%
    \hbox to \linewidth{%
      \raisebox{0.6ex}{\rule{1.5em}{0.4pt}}%
      \hspace{0.5em}%
      {\bfseries #1}%
      \hspace{0.5em}%
      \leaders\hbox{%
        \raisebox{0.6ex}{\rule{1pt}{0.4pt}}%
      }\hfill
    }%
  }%
}
\graphicspath{{./figures/}}

\newtcolorbox{userquery}{colback=gray!8, colframe=scipostdeepblue,
    fonttitle=\bfseries\color{white}, title=User, boxrule=0.6pt, arc=2pt,
    left=4pt, right=4pt, top=3pt, bottom=3pt}

\begin{document}

\begin{center}
    {\Large\textbf{\color{scipostdeepblue}{Agentic Re-Casting using Agentic Re-Simulations}}}
\end{center}

\begin{center}\textbf{
    Sascha~Diefenbacher\textsuperscript{1},
    Tilman~Plehn\textsuperscript{1,2},
    Daniel~Schiller\textsuperscript{1}, and
    Nikita~Schmal\textsuperscript{1}
}\end{center}

\begin{center}
    {\bf 1} Institut f\"{u}r Theoretische Physik, Universit\"{a}t Heidelberg, Germany\\
    {\bf 2} Interdisciplinary Center for Scientific Computing (IWR), Universit\"{a}t Heidelberg, Germany\\
\end{center}

\begin{center}
    \today
\end{center}

\section*{\color{scipostdeepblue}{Abstract}}
\textbf{\boldmath{Analysis re-casting at the LHC is highly standardized and nevertheless requires resources, time, and physics input. Building on the new MadAgents.v3, we show how a global SFitter analysis can be updated by an agentic system with a physicist in the loop. The agentic interface allows us to make the advanced SFitter methodology available to a wider audience. All physical and technical aspects of this agentic re-casting study can be trivially generalized beyond SFitter.}}

{
\vspace{5pt}
\noindent \url{https://github.com/heidelberg-hepml/SFitterAgents}
}

\vspace{10pt}
\noindent\rule{\textwidth}{1pt}
\tableofcontents\thispagestyle{fancy}
\noindent\rule{\textwidth}{1pt}
\vspace{10pt}

\clearpage
\section{Introduction}
\label{sec:intro}

The HL-LHC program will continue the fundamental paradigm shift towards data-driven precision analyses to answer the question \textsl{What Lagrangian describes the LHC data altogether?} The most suitable theoretical interpretation framework is SMEFT, as it covers extensions of all sectors of the Standard Model, combines many rate and kinematic measurements in a meaningful manner, and can be evolved across relevant energy scales. In the Higgs-gauge sector~\cite{Corbett:2012ja,Butter:2016cvz,Ellis:2018gqa,Biekotter:2018ohn,Kraml:2019sis,Almeida:2021asy}, the top sector~\cite{Buckley:2015lku,Hartland:2019bjb,Brivio:2019ius,Aoude:2022deh}, their combination~\cite{Ellis:2020unq,Ethier:2021bye,Elmer:2023wtr,Celada:2024mcf}, and including flavor physics~\cite{Bissmann:2020mfi,Bruggisser:2022rhb,Bartocci:2023nvp}, such global SMEFT analyses have a long tradition. 

A critical aspect of global analyses is the re-interpretation of published measurements in terms of novel theory hypotheses~\cite{DPHEP:2025prk}. Only few experimental analyses test a range of theory hypotheses affecting the same kinematic regime~\cite{ATLAS:2025oiy}. Global analyses therefore replace the original theory hypothesis with SMEFT. While public likelihoods simplify such a re-interpretation~\cite{Elmer:2023wtr}, implementing measurements re-analyzing them under the new theory hypothesis takes serious effort. In \sfitter, including a new analysis takes months, providing a formidable benchmark for workflow acceleration through agentic AI. 

Whereas proper LHC data analysis is a highly non-trivial task with many pitfalls in a highly specialized statistical framework, analysis re-casting largely relies on simulation (\mg) and a well-defined likelihood analysis (\sfitter). For the simulation, \ma~\cite{Plehn:2026gxv} already provide an agentic interface that covers code installation, user training, experienced user support, all the way to autonomous simulation campaigns. In recent months, there have been many ideas how analyses~\cite{Diefenbacher:2025zzn,Gendreau-Distler:2025fsj,Esmail:2026jpb,Birk:2026zpd,Desai:2026nmx}, data interpretation~\cite{Bakshi:2025fgx,Menzo:2025cim,Qiu:2026iby,Agrawal:2026lvg,Faroughy:2026dkj,Costa:2026oew,Wang:2026jjn,Saad:2026pan}, and theory~\cite{Menzo:2026qrl,Niarchos:2026fbt} for the LHC and beyond can be accelerated by agentic tools. 

We first introduce a new version of \ma, suitable for LHC analyses and analysis re-casting. The original \mav{1} uses OpenAI API keys, while \mav{2} was migrated to Claude, largely by Claude, with the help of Siddharth Mishra-Sharma. Both are documented in Ref.~\cite{Plehn:2026gxv}. The new \mav{3} is inspired by a technical workshop with \href{https://labs.barrabytes.com}{barra labs} and focuses on trustworthy simulation support. Work on an exciting \mav{4} will begin the week after this paper has been submitted to the arXiv, illustrating the pace of progress in scientific AI.\footnote{Agentic LLMs are transforming particle physics research, but it is not clear if using commercial or public LLMs and expert advice from industry to accelerate physics workflows should be considered physics research. This is why we submit \sfa to SciPost Physics Codebases.}

In parallel, we also introduce \sfa, an agentic framework that allows a broad audience to work with \sfitter and accelerates the work of expert users. We design \sfa as an interactive research partner, rather than as a fully automated system. At every step, key decisions about the underlying physics of global SMEFT analyses are driven by the physicist. As a result, the \sfa are specialized in design and orchestration of simulation campaigns, data visualization and validation, and the in-depth operational details of \sfitter. The agents have enough understanding of global analyses to provide input on each action, without taking control away from the physicist.

In Sec.~\ref{sec:sf} we introduce \sfitter, its defining uncertainty budget, the structure of its likelihood, and the underlying SMEFT hypothesis with the Run~2 top-sector dataset~\cite{Elmer:2023wtr}. This dataset requires updating, defining the \sfa task. In Sec.~\ref{sec:setup} we present the new \mav{3}. We then develop the agentic interface to analysis re-interpretation in general and \sfitter in particular. We employ these \sfa in Sec.~\ref{sec:end_to_end}, to show how they can accelerate the standard workflow and support the student or expert in the loop. We show how step by step \sfa can add a new kinematic $t\bar{t}Z$ measurement to the global SMEFT analysis, without and with published likelihoods. In the appendices we provide a validation using silent or hard-to-catch failure modes and a repeatable or reproducible benchmark.

\section{SFitter global analyses and re-casting}
\label{sec:sf}

Naively, global LHC analyses would combine measurements from different signatures and fiducial regions, where the theory hypothesis of all original analyses is the same as that of the global analysis. In reality, this would limit the scope and impact of global analyses. An inspiring example is a global SMEFT analysis incorporating resonance searches for heavy gauge bosons $V' \to VH$. Their final state is two fat jets and we cannot really isolate the SM $VH$ background in this final state. At large $m_{VH}$ we can nevertheless search for dimension-6 contributions to $VH$ production. An \sfitter re-casting and global analysis then proceeds in consecutive steps:
\begin{enumerate}
\item Extraction of a measurement from an experimental publication;
\item SM signal and background re-simulation of the published analysis;
\item New SMEFT simulation for the new signal hypothesis;
\item Global \sfitter likelihood analysis.
\end{enumerate}

\subsection{SFitter motivation}
\label{sec:sf-stucture}

As this publication combines a first public release of \sfitter with an accompanying agentic framework, we briefly review the target and structure of \sfitter global analyses. \sfitter was originally developed for a global analysis of supersymmetric signatures at the LHC~\cite{Lafaye:2004cn,Lafaye:2007vs}. Three technical aspects still define the current applications: $(i)$ a combination of total cross section and kinematic measurements; $(ii)$ a comprehensive uncertainty treatment, including theory uncertainties; $(iii)$ a flexible choice of marginalization and profiling of the fully exclusive likelihood, in terms of parameters of interest and nuisance parameters.

Since supersymmetric particles are still waiting to be discovered at the LHC, \sfitter was extended to global Higgs analyses~\cite{Lafaye:2009vr} and first applied to data for a global analysis of Higgs couplings~\cite{Klute:2012pu}. In the Higgs sector, it is important to treat degenerate local likelihood maxima arising from sign changes in the effective couplings. Technically, these early applications introduced weighted and cooling Markov chains for an efficient numerical construction of the fully exclusive likelihood. From a statistics perspective, the fully exclusive \sfitter likelihood can be treated in a frequentist approach, but in some analyses we resort to the Bayesian concept of toy measurements~\cite{Brivio:2022hrb,Costantini:2024wby}.

The current SMEFT interpretation framework of global \sfitter analyses arises naturally when combining Higgs physics with the physics of massive weak bosons~\cite{Butter:2016cvz}. A range of dimension-6 operators can be probed in $VH$ production as well as in $VV$ production, so complementing the Higgs sector with di-boson production significantly boosts the reach of the global analysis~\cite{Biekotter:2018ohn}. For all SMEFT analyses, \sfitter includes not only the leading interference between the SM and the dimension-6 contribution, but also the dimension-6 squared contribution. Higher-dimension operators can, in principle, be included~\cite{Dawson:2021xei,Dawson:2024ozw}, but a systematic treatment of all dimension-8 operators would exceed the predictive power of the current LHC dataset and challenge the numerical construction of the likelihood. An interesting and relevant question is to what degree global analyses truncated to dimension-6 operators in SMEFT differ from an \sfitter analysis of the underlying UV-complete model~\cite{Brivio:2021alv}.

To systematically include the top sector in the global \sfitter analysis we need to account for higher-order QCD predictions and their theory uncertainty~\cite{Hocker:2001xe}. For the Higgs sector this information is readily available from the Higgs cross section group~\cite{LHCHiggsCrossSectionWorkingGroup:2011wcg,Dittmaier:2012vm}. For the top-sector \sfitter analysis~\cite{Brivio:2019ius} we rely heavily on NLO predictions from \mg. This increases the scope of the simulation campaigns significantly, a problem that by now can be solved with \ma.

In the classical setup of global analyses, the availability of detailed information about experimental analyses is a bottleneck, and reverse-engineering is extremely time-consuming. Recently, ATLAS and CMS have started to provide public likelihoods for their analyses, which allow us to reproduce their published results to the line width in their figures~\cite{Elmer:2023wtr}. The availability of information will play an important role as a requirement for the \sfitter agents. Finally, the most recent technical improvements of \sfitter employ neural importance sampling to construct and evaluate the exclusive likelihood much more efficiently~\cite{Heimel:2024drk} and show how unbinned simulation-based inference can benefit global SMEFT analyses~\cite{Bahl:2025mib}.

\sfitter includes a link to renormalization group evolution for SUSY models~\cite{Adam:2010uz} and flavor physics~\cite{Bartocci:2024fmm}. In addition, the \sfitter methodology has been applied beyond LHC physics, for instance to global dark matter analyses~\cite{Butter:2016tjc}, to the link between hard LHC scattering and flavor physics~\cite{Bruggisser:2021duo,Bartocci:2023nvp}, the link to axions~\cite{Biekotter:2023mpd}, and a global analysis of electric dipole moment measurements at the GeV scale~\cite{Degenkolb:2024eve}. This broad range of applications over more than 20 years of \sfitter and the continuous technical improvements of the methods and the implementation is the main motivation for the agent-supported public version of \sfitter.

\subsection{SFitter likelihood}
\label{sec:sf-like}

To construct the \sfitter likelihood as a function of the model parameters $c$ for a single counting measurement we begin by taking the Poisson probabilities to observe $d$ events with $m$ events predicted, multiplied with the probability to measure the number of background events in the signal region $b$ compared to those in the control region $b_\text{CR}$, using an appropriate interpolation factor $k$~\cite{Lafaye:2004cn}
\begin{align}
  L_\text{excl}(c, \theta, b)
  = \pois(d|m(c, \theta, b)) \;
  \pois(b_\text{CR}|bk)\prod_i \,
  \mathcal{C}_i(\theta_i, \sigma_i) \; .
  \label{eqn:Excl_Likeli}
\end{align}
The distribution of the nuisance parameters $\theta$ is constrained by the $\mathcal{C}_i$ terms. The type of constraint depends on the uncertainty considered. Experimental systematics, often results of an auxiliary measurement, are modeled as Gaussians $\mathcal{N}$ with mean $\theta_i$ and width $\sigma_i$. In the case of theory uncertainties we decide to model them as flat likelihoods~\cite{Hocker:2001xe,Ghosh:2022lrf}
\begin{align}
    \mathcal{F}(x|\theta, \sigma) = \frac{1}{2\sigma}\Theta[x - (\theta - \sigma)]\Theta[(\theta + \sigma) - x] \, ,
\end{align}
with half-width $\sigma$ around the central value $\theta$.

A key aspect of the \sfitter likelihood is that all systematic uncertainties of the same type are fully correlated. This is done through a correlation matrix between measurements
\begin{align}
    C_{ij} = \frac{\sum_{\text{syst}} \rho_{ij} \sigma_{i,\text{syst}}\sigma_{j,\text{syst}} }{\sigma_{i,\text{exp}}\sigma_{j,\text{exp}}} 
    \qquad  \text{with} \qquad 
    \sigma_{i,\text{exp}}^{2} = \sum_{\text{syst}}\sigma_{i,\text{syst}}^{2} + \sum_{\pois}\sigma_{i,\pois}^{2} \, ,
\end{align}
where the indices $i,j$ run over all measurements, and $\rho_{ij} = 0.99$ ensures the invertibility of the correlation matrix.

\subsubsection*{Higgs-gauge sector}

The Higgs and di-boson dataset does in general not allow for the Gaussian approximation to hold, due to small signals in the presence of large backgrounds. The exclusive likelihood in Eq.\eqref{eqn:Excl_Likeli} cannot be simplified, and we need to profile over Poisson, Gaussian, and flat distributions.  If we are interested in the observed signal number $s = d - b$, we write the Poisson terms as a function of $\tilde{s} = m - b$
\begin{align}
    P(b,d|\tilde{s},\tilde{b}) = \frac{e^{-(\tilde{s}+\tilde{b})}(\tilde{s}+\tilde{b})^{d}}{d!}\frac{e^{-\tilde{b}}(\tilde{b})^b}{b!} \; .
\end{align}
We then define a generalized $\chi^{2}$, which vanishes when data and prediction match, 
\begin{align}
    \chi^{2} &= -2 \log 
    \frac{P(b_\text{CR},d|\tilde{s},\tilde{b})}{P(\tilde{b},\tilde{d}|\tilde{s},\tilde{b})} \notag  \\
    &= -2 \max_{\tilde{b}} \, \left[ (d-(\tilde{s}+\tilde{b}))\log(\tilde{s} + \tilde{b}) + (b_\text{CR} - \tilde{b})\log(\tilde{b}) + \log\left(\frac{(\tilde{s}+\tilde{b})!}{d!}\frac{\tilde{b}!}{b_\text{CR}!}\right) \right] \; .
\end{align}
Next, we profile over the expected background $\tilde{b}$. Doing this for each data point would be inefficient, so we approximate this contribution by splitting it into two parts
\begin{align}
    \log \, L_{\pois,d}(\Tilde{s}|d,b_\text{CR}) 
    &= (d - (\tilde{s}_{\sigma} 
    + b_\text{CR})) \log (\tilde{s}_{\sigma} + b_\text{CR}) 
    + \log \frac{(\tilde{s}_{\sigma}+b_\text{CR})!}{d!} \notag \\
    \log \, L_{\pois,b}(\Tilde{s}|d,b_\text{CR}) 
    &= (b_\text{CR} - (d - \tilde{s}_{\sigma})) \log (d - \tilde{s}_{\sigma}) 
    + \log \frac{(d - \tilde{s}_{\sigma})!}{b_\text{CR}!} \; .
\end{align}
We account for flat nuisance parameters by introducing the shifted signal $\tilde{s}_{\sigma} = \tilde{s} \pm \sigma_\text{theo}$, where the sign is chosen such that the signal is shifted towards data. The last missing piece is the Gaussian systematics. In analogy to the signal strengths, they can be computed as
\begin{align}
    -2 \, \log \, L_\text{Gauss}(\Tilde{s}|d,b_\text{CR}) 
    = \frac{(d - b_\text{CR} - \tilde{s}_{\sigma})^{2}}{\sum_\text{syst} (\sigma_{d,i} - \sigma_{b,i})^{2}} \, .
\end{align}
Finally, we combine all contributions using the approximate formula~\cite{Lafaye:2009vr}
\begin{align}
    \frac{1}{\log L_\text{full}} \approx \frac{1}{\log L_\text{Gauss}} + \frac{1}{\log L_{\text{Poiss},b}} + \frac{1}{\log L_{\text{Poiss},d}} \, ,
\end{align}
which is exact in the Gaussian limit.

\subsubsection*{Top sector}

Profiling over nuisance parameters is particularly easy for the top dataset, which consists of unfolded data and signal strengths with negligible backgrounds,
\begin{align}
  L_\text{prof}(c) = \max_\theta \ \pois(d|m(c, \theta))
  \prod_i \, \mathcal{N}(\theta_{i},\sigma_{i}) \;
  \prod_j \, \mathcal{F}(\theta_{j},\sigma_{j}) \; .
\end{align}
All signals are large enough to approximate the Poisson distribution by a Gaussian. The convolution of Gaussians becomes a single Gaussian with total width $\sigma_\text{syst}^{2} = \sum_{i} \sigma_{syst, i}^{2}$. The flat theory uncertainties lead to a shift of the theory prediction by $\sigma_\text{theo} = \sum_{i}\sigma_{theo,i}$ towards the data. By profiling, uncorrelated theory uncertainties add linearly, while systematics add in quadrature. The likelihood in the Gaussian limit then becomes
\begin{align}
    \sqrt{ -2 \, \log L_\text{prof}(c)} = 
    \begin{cases}
        (d - ( m + \sigma_{\text{theo}}))/\sigma_\text{syst} & d > m + \sigma_\text{theo}\\
        0 & d \in [m-\sigma_\text{theo}, m + \sigma_\text{theo}]\\
        (d - (m -  \sigma_\text{theo}))/\sigma_\text{syst} & d < m - \sigma_\text{theo} \; ,
    \end{cases}
\end{align}
where we absorb the normalization constants such that $L_\text{prof} = 1$ when the data falls within the theory uncertainty.

\subsection{Top-SMEFT dataset}
\label{sec:sf-data}

\begin{table}[b!]
    \centering
    \begin{small} \begin{tabularx}{\textwidth}{lXlX}
    \toprule
        Operator & Definition & Operator & Definition \\
        \midrule
        $\ope_{Qq}^{1,8}$ & $(\bar{Q}\gamma_\mu T^A Q) \; (\bar{q}_i\gamma^\mu T^A q_i)$ &
        $\ope_{tu}^8$ & $(\bar{t}\gamma_\mu T^A t) \; (\bar{u}_i\gamma^\mu T^A u_i)$ \\
        $\ope_{Qq}^{1,1}$ & $(\bar{Q}\gamma_\mu Q) \; (\bar{q}_i\gamma^\mu q_i)$ &
        $\ope_{tu}^1$ & $(\bar{t}\gamma_\mu t) \; (\bar{u}_i\gamma^\mu u_i)$ \\
        $\ope_{Qq}^{3,8}$ & $(\bar{Q}\gamma_\mu T^A\tau^I Q) \; (\bar{q}_i\gamma^\mu T^A \tau^I q_i)$ &
        $\ope_{td}^8$ & $(\bar{t}\gamma^\mu T^A t) \; (\bar{d}_i\gamma_\mu T^A d_i)$ \\
        $\ope_{Qq}^{3,1}$ & $(\bar{Q}\gamma_\mu\tau^I Q) \; (\bar{q}_i\gamma^\mu\tau^I q_i)$ &
        $\ope_{td}^1$ & $(\bar{t}\gamma^\mu t) \; (\bar{d}_i\gamma_\mu d_i)$ \\
        \midrule
        $\ope_{Qu}^8$ & $(\bar{Q}\gamma^\mu T^A Q) \; (\bar{u}_i\gamma_\mu T^A u_i)$ &
        $\ope_{Qd}^1$ & $(\bar{Q}\gamma^\mu Q) \; (\bar{d}_i\gamma_\mu d_i)$ \\
        $\ope_{Qu}^1$ & $(\bar{Q}\gamma^\mu Q) \; (\bar{u}_i\gamma_\mu u_i)$ &
        $\ope_{tq}^8$ & $(\bar{q}_i\gamma^\mu T^A q_i) \; (\bar{t}\gamma_\mu T^A t)$ \\
        $\ope_{Qd}^8$ & $(\bar{Q}\gamma^\mu T^A Q) \; (\bar{d}_i\gamma_\mu T^A d_i)$ &
        $\ope_{tq}^1$ & $(\bar{q}_i\gamma^\mu q_i) \; (\bar{t}\gamma_\mu t)$ \\
        \midrule
        $\ope_{\phi Q}^{1}$ & $(\phi^\dagger\,i \stackrel{\longleftrightarrow}{D_\mu} \phi) \; (\bar{Q}\gamma^{\mu}Q)$ &
        $^\ddagger \ope_{tB}$ & $(\bar{Q}\sigma^{\mu\nu} t)\,\widetilde{\phi}\,B_{\mu\nu}$ \\
        $\ope_{\phi Q}^{3}$ & $(\phi^\dagger\,i \stackrel{\longleftrightarrow}{D_\mu^I} \phi) \; (\bar{Q}\gamma^{\mu}\tau^I Q)$ &
        $^\ddagger \ope_{tW}$ & $(\bar{Q}\sigma^{\mu\nu} t)\,\tau^I\widetilde{\phi}\,W_{\mu\nu}^I$ \\
        $\ope_{\phi t}$ & $(\phi^\dagger\,i \stackrel{\longleftrightarrow}{D_\mu} \phi) \; (\bar{t}\gamma^{\mu}t)$ &
        $^\ddagger \ope_{bW}$ & $(\bar{Q}\sigma^{\mu\nu} b)\,\tau^I\phi \,W_{\mu\nu}^I$ \\
        $^\ddagger \ope_{\phi tb}$ & $(\widetilde{\phi}^\dagger iD_\mu \phi) \; (\bar{t}\gamma^{\mu}b)$ &
        $^\ddagger \ope_{tG}$ & $(\bar{Q}\sigma^{\mu\nu} T^A t)\,\widetilde{\phi}\,G_{\mu\nu}^A$ \\
        \bottomrule
    \end{tabularx} \end{small}
    \caption{22 independent top-sector operators, related to the Warsaw basis in the Appendix of Ref.~\cite{Brivio:2019ius}.}
    \label{tab:top_operators}
\end{table}

The immediate goal of this paper is to extend the global SMEFT analysis in the top sector from Refs.~\cite{Heimel:2024drk, Brivio:2019ius}. We show that the agentic system reproduces our previous results and extends them using new measurements. Eventually, this analysis should be updated with Run~3 data, and we will show that this should be done with full agentic support. 

We parametrize the effect from higher dimension operators $\ope$ using Wilson coefficients $C$ within the SMEFT Lagrangian
\begin{align}
    \mathcal{L}_\text{SMEFT} = \mathcal{L}_\text{SM} + \sum_k\left(\frac{C_k}{\Lambda^2} {^\ddagger}\ope_k + \text{h.c.} \right) + \sum_l\frac{C_l}{\Lambda^2}\ope_l \, .
\end{align}
The operator basis is given in Table~\ref{tab:top_operators}, with non-hermitian operators denoted as $^\ddagger \ope$. We impose a $U(2)$ flavor symmetry on the first and second quark generations, as most top observables are blind to the flavor of light quarks,
\begin{alignat}{5}
q_i & = (u^i_L,d^i_L) & \qqquad 
u_i & = u^i_R,\,d_i=d^i_R 
\quad \text{for} \quad i=1,2 \notag  \\
Q & = (t_L,b_L) & \qqquad 
t &= t_R,\,b=b_R \; .
\label{eq:symm}
\end{alignat}
All quark masses except for the top mass are taken to be zero, which leaves 22 independent operators.

These top-sector operators fall into three categories: $(i)$ the 4-fermion currents with $RR$ and $LL$ helicity structure; $(ii)$ operators with $RL$ or $LR$ helicity flips; $(iii)$ couplings of heavy quarks to gauge bosons. We apply additional relations arising from gauge invariance
\begin{align}
C_{\phi Q}^- = C_{\phi Q}^1 - C_{\phi Q}^3 
\qquad \text{and} \qquad
C_{tZ} = c_w C_{tW} - s_w C_{tB} \; ,
\end{align}
and choose $C_{\phi Q}^{-},C_{\phi Q}^{3},C_{tW}$ and $C_{tZ}$ as our independent model parameters. 

The first two kinds of operators are mainly constrained by observables involving top pair production, such as $t\bar{t}$ and associated $t\bar{t}W, t\bar{t}Z$ final states. To fully constrain the remaining operators, we also include single top production, associated $tZ$ and $tW$ production, and top decay observables. The measurements entering our global analysis are listed in App.~\ref{app:top_data}. A detailed description of the measurements and the impact of the different Wilson coefficients is given in Ref.~\cite{Elmer:2023wtr}. 

\section{Agents with agents}
\label{sec:setup}

\begin{figure}[b!]
    \includegraphics[width=\linewidth]{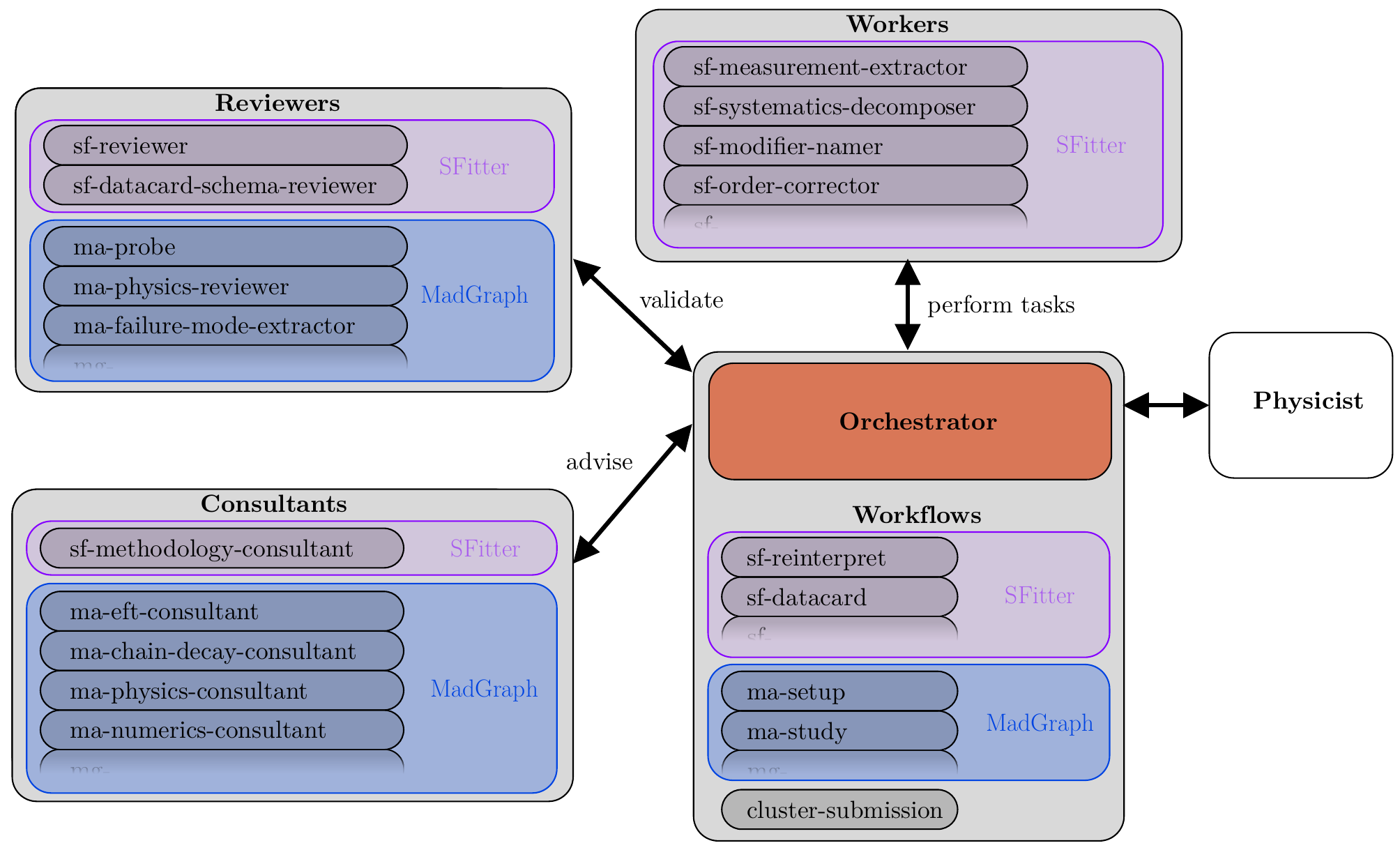}
    \caption{\sfa structure.}
    \label{fig:sfa_scheme}
\end{figure}

LLMs for physics research have a rich history, starting with astrophysical and cosmological applications. A first obvious LLM task is literature search, for instance using the non-generative astroBERT~\cite{grezes2021buildingastrobertlanguagemodel} and the generative  AstroLLaMA~\cite{Nguyen:2023nhp}.

Processing relevant aspects of physics data using LLMs is much less established. L3M~\cite{Heneka:2025fpe} showed how pre-trained LLMs can be fine-tuned to process SKA light cones with the help of simple connector networks. Recently, we have shown how LLMs can be used to condition LHC event generators on Feynman diagrams, allowing zero-shot event generation beyond the training processes and significantly accelerating the generator training~\cite{Bahl:2026lsa}.

Agentic LLM applications have a rapidly growing field. The first agents for experimental LHC physics were Agents for Discovery~\cite{Diefenbacher:2025zzn}, automatizing anomaly searches. For cosmological simulation tools, LLM agents provide help to set up, run, and analyze the simulations, for instance, for CLASS/\href{https://github.com/santiagocasas/clapp/}{CLAP} and \href{https://github.com/franciscovillaescusa/CAMELS_Agents/}{CAMEL}. In LHC physics, the first agents that support a standard simulation tool are \ma~\cite{Plehn:2026gxv}. We emphasize that none of the currently available fundamental physics agents do their own analysis or simulation. They `only' provide extremely efficient interfaces to domain-specific and highly advanced analysis or simulation tools. 

We develop \sfa, using an updated version of \ma~\cite{Plehn:2026gxv}, to support re-simulation and global analyses. We integrate the \sfitter tool, such that the same agentic system accompanies a set of measurements from the first corresponding simulations to the global SMEFT constraint. The agent structure is illustrated in Figure~\ref{fig:sfa_scheme}. The worker-based \sfa follow the original \ma structure, while \mav{3} employs consultants. Here, the orchestrator can use the in-depth knowledge of the consultants, which do not carry out the work themselves but only provide information. 

\subsection{\mav{3}}

Our updated \mav{3} with consultant agents is built on the premise that a trustworthy \mg simulation should not rely on a public LLM alone. Relevant details are often highly specific, ranging from command syntax and default settings to conventions and internal behavior. Errors that prevent a simulation from running provide explicit feedback, allowing an agent to iteratively correct the setup. More problematic are configurations that execute successfully but do not correspond to the user's intended simulation. This so-called silent failure mode does not produce well-defined error messages, so we need to design our updated architecture and operating principles to prevent it.

\begin{table}[t]
\centering
\begin{small} \begin{tabularx}{\linewidth}{@{}lX@{}}
    \toprule
    Group                      & Subagents \\
    \midrule
    Process specification (LO) & process-syntax-, chain-decay-, coupling-order-, diagram-filter-, diagram-enumeration-, polarization-consultant \\
    Models                     & model-loader-, ufo-, restriction-, nlo-model-, eft-consultant \\
    NLO                        & nlo-syntax-, amcatnlo-, nlo-export-, madloop-, fks-consultant \\
    Decays and widths          & madwidth-, madspin-interface-, bw-window-consultant \\
    Amplitude and output       & helas-amplitude-, aloha-, color-decomposition-, output-consultant \\
    Run cards and kinematics   & param-card-, scales-pdf-, kinematic-cuts-consultant \\
    Runtime and integration    & phase-space-, mc-integration-, launch-, matching-, systematics-consultant \\
    External-tool interfaces   & pythia8-interface-, delphes-interface-, madanalysis5-interface-, rivet-interface-, interface-consultant \\
    Installation               & installation-consultant \\
    \midrule
    Reasoning domains          & physics-, math-, numerics-consultant \\
    \midrule
    Reviewers                  & physics-, math-, numerics-reviewer \\
    Auxiliary verification     & probe, blind-spot-auditor, failure-mode-extractor \\
    \bottomrule
\end{tabularx} \end{small}
\caption{\mav{3} roster, grouped into \mg consultants (top), reasoning consultants (center), and reviewing subagents (bottom).}
\label{tab:ma-roster}
\end{table}

\ma remain an orchestrating architecture, detailed in Table~\ref{tab:ma-roster}. The orchestrator coordinates a roster of around forty specialized subagents, each responsible for a distinct part of the simulation, from process syntax and model loading to NLO computations, decays, run cards, and interfaces to external tools. Three additional subagents provide physics, mathematical, and numerical reasoning. Each subagent contributes detailed knowledge within its domain and rejects queries outside it. The orchestrator routes each question to the responsible subagent and assembles their returns into an answer, rather than answering from its own \mg knowledge. Reviewers check the assembled response.

We adopt the consultant-based system in response to the increasing capabilities of modern LLMs as general-purpose models. Routing generic tasks to separate worker agents, such as a general coding agent, introduces additional token and inference overhead without necessarily providing a corresponding benefit. More importantly, it can reduce performance if the orchestrator fails to propagate all relevant context and instructions to the worker agent. We therefore let the orchestrator handle generic tasks directly, with the consultants performing an information retrieval tasks.

The orchestrator and the subagents share a number of operating principles, governing how to establish what is true, decide when a result can be trusted, and improve with use:
\begin{enumerate}

\item \textbf{Source grounding \quad}
Every \mg-specific claim, such as an accepted syntax or a default value, must be grounded in the locally installed \mg. This can be done either by inspecting the source code or by running a dedicated runtime check. Such claims may never be taken from documentation. Knowledge from the LLM weights provides an intuition but must be verified locally. This source grounding is the foundation of the updated \ma.

\item \textbf{Lasting records \quad}
The orchestrator and each consultant keep track of their knowledge in the style of \href{https://gist.github.com/karpathy/442a6bf555914893e9891c11519de94f}{llm-wiki}. The orchestrator records the subagents required for a task and in what order to call them. Consultants record \mg facts affecting their task and check them against the source. Each agent always has short descriptions of its records in context, allowing it to retrieve the full record when needed.

Whenever an agent considers something worth keeping while working on a task, it becomes part of the record. In addition, records can be written ahead of a task by studying \mg or by turning a past mistake into a lesson. A new setup is expensive, since the knowledge must be extracted from the \mg source, derived, and checked. Once recorded, subsequent tasks rely on the cheap records. To keep their records from growing too large, the agents maintain them.

\item \textbf{Completion vs correctness \quad}
A completed run without error messages does not guarantee that its physics is correct. The agent defines its own checks to separate a correct setup from a wrong one, such as a branching ratio staying below one, a non-zero cross section, or a set of Feynman diagrams for a process. To stabilize that target, the agent keeps the user-specified physics apart from the simulation outcome. Reconciling them flags setups that do not compute the desired quantity.

\item \textbf{Recorded confidence \quad}
Not all consultant claims are equally certain. Some are read from the source, while others require a long chain of reasoning with potentially wrong steps. Some might even be wild guesses. Recorded claims do not carry information on the reasoning behind them, so they have to be marked by how directly the source supports them. The orchestrator verifies or openly hedges whatever is not grounded, so a confident answer always rests on cited evidence.

\item \textbf{Adversarial review \quad}
On request, a finished \mg setup can be reviewed, with a focus on failure modes. The review starts from the claims the consultants made when building the setup. The orchestrator calls each consultant back in a fresh context, so they cannot simply repeat earlier reasoning. If a claim can be tested by running \mg, the output is compared with physics expectations, like the order of magnitude of the cross section. If a claim can only be reasoned, it is re-derived from the \mg source, or checked by the physics, mathematics, or numerics reviewer.
\end{enumerate}
In addition, the orchestrator commands a small set of workflows for recurring multi-step tasks. They build and reconcile a setup, run a deeper check on request, study \mg ahead of a task, and harden the system after a mistake. The orchestrator may follow these pre-defined workflows whenever they fit. We validate the updated \ma in detail in App.~\ref{app:silent-failure-benchmark}.

\subsection{\sfa}

\begin{table}[t!]
\centering
\begin{small} \begin{tabularx}{\linewidth}{@{}lX@{}}
    \toprule
    Subagent                      & Task \\
    \midrule
    \textbf{Consultant}           & \\
    sf-methodology-consultant     & Advises on the \sfitter likelihood framework, such as which distribution each uncertainty enters, the correlation structure, and the global analysis settings. \\
    \midrule
    \textbf{Workers}              & \\
    \catrule{Measurement extraction} \\ \addlinespace[2pt]
    sf-measurement-extractor      & Retrieves the per-bin measurement and its metadata from HEPData or the publication, and classifies it as absolute or normalized and fiducial or total. \\
    sf-systematics-decomposer     & Splits the total systematic uncertainty into per-bin, per-source contributions. \\
    sf-modifier-namer             & Maps each uncertainty source onto \sfitter's canonical modifier name and correlation label. \\
    \catrule{SM-prediction template} \\ \addlinespace[2pt]
    sf-order-corrector            & Lifts the \mg baseline to the required perturbative order with the per-bin NNLO/NLO $K$-factor. \\
    sf-scale-uncertainty-estimator & Estimates the renormalization and factorization scale uncertainty on the prediction template. \\
    sf-pdf-uncertainty-estimator  & Estimates the PDF and $\alpha_s$ uncertainty on the template. \\
    sf-parametric-uncertainty-estimator & Estimates the remaining template uncertainties from the top mass, parton shower, MC statistics, and EFT truncation. \\
    \catrule{EFT parametrization} \\ \addlinespace[2pt]
    sf-scan-designer              & Designs the Wilson-coefficient scan that feeds the EFT parametrization. \\
    sf-scan-sanitizer             & Zeroes the SMEFTatNLO param card so no unscanned operator contaminates the scan. \\
    sf-kappa-extractor            & Extracts the per-bin dependence on the Wilson coefficients, both diagonal and interference, from the scan output. \\
    sf-convention-translator      & Translates the operator output into \sfitter's naming and normalization convention. \\
    \catrule{Datacard assembly} \\ \addlinespace[2pt]
    sf-id-encoder                 & Builds and validates the \sfitter measurement identifier. \\
    \midrule
    \textbf{Reviewers}             & \\
    sf-reviewer                   & Reviews the whole reinterpretation for faults, such as broken extraction traceability, the wrong SM order, or scan contamination. \\
    sf-datacard-schema-reviewer   & Validates the assembled datacard against the schema and the datacard parser. \\
    \bottomrule
\end{tabularx} \end{small}
\caption{\sfa roster, building on the \ma in Table~\ref{tab:ma-roster} and grouped into an \sfitter methodology consultant (top), workers (center), and reviewers (bottom).}
\label{tab:sf-roster}
\end{table}

While \mg is a large and complex software framework requiring many topic-specific consultants, the published part of \sfitter is comparatively simple. Its challenges are internal conventions and dedicated workflows, which the \sfa have to follow. We list the three types of subagents in Table~\ref{tab:sf-roster}: one consultant advises on the \sfitter methodology, a dozen workers carry out the analysis, from extracting a measurement to assembling the datacard, and two reviewers check their results.

The \sfa extend the \ma as illustrated in Figure~\ref{fig:sfa_scheme}, naturally following the same operating principles. Software-specific facts are verified against the installed \sfitter code, while conventions and workflows are extracted from a curated documentation. The documented record covers the measurement extraction, the theory uncertainty, the SMEFT operator basis, the datacard format, and the statistical methodology.

The \sfa follow the standard steps of (re-)interpreting a measurement: $(i)$ extract the measurement from the publication, its per-bin values and uncertainties; $(ii)$ build the Standard-Model simulation and its theory uncertainty budget; $(iii)$ define and simulate the SMEFT hypothesis, scan the Wilson coefficients, and extract the dependence of each bin on the Wilson coefficients; $(iv)$ assemble the new datacard, validate it, and run the statistical analysis. Each step is covered by a set of workers listed in Table~\ref{tab:sf-roster}. The second and third steps also use \ma for the (re-)simulation. Each finished step is reviewed before the next step begins. The user can intervene at any step, skip a step, or iterate it.

\section{Agentic global analysis and re-casting}
\label{sec:end_to_end}

Following the re-casting workflow introduced in Sec.~\ref{sec:sf} we now show how the \sfa follow these steps, always talking to a physicist. We do not build the kind of autonomous \ma campaign shown in Ref.~\cite{Plehn:2026gxv}, because we want to automize the standard re-casting steps while giving the physicist the opportunity to steer the re-analysis in detail.

\subsection{Measurement extraction}

First, we need to find a measurement to implement, with the expectation that it improves the global analysis. We ask the \sfa to run a global SMEFT analysis using the already existing top dataset~\cite{Heimel:2024wph} and then ask them to study the constraints and suggest a new measurement to include:
\begin{userblock}[User]
    My job is to extend the currently implemented top dataset to improve the global top SMEFT analysis. Can you perform a quick run using the current dataset to check the constraints? Using these give me suggestions on what type of measurement to add to improve our analysis. As a final deliverable I would like to have a list of different measurements sorted by priority.
\end{userblock}
\begin{promptblock}[\sfa (shortened)]
Prioritized list of measurements to add, ordered by constraining-power gain.
\begin{itemize}
    \item Differential $t\bar{t}Z$, $\mathrm{d}\sigma/\mathrm{d}p_{T,Z}$, together with $t\bar{t}Z/tZ$ polarization and angular observables. This targets the weakest-constrained directions,
    $\mathcal{O}_{\phi t}$, $\mathcal{O}_{\phi q}^{-}$, $\mathcal{O}_{tZ}$, and
    $\mathcal{O}_{\phi q}^{3}$, with polarization observables required to break the
    $\mathcal{O}_{\phi t}\leftrightarrow\mathcal{O}_{\phi q}^{-}$ degeneracy.

    \item $t\bar{t}$ charge asymmetry $A_C$ together with boosted high-mass
    $t\bar{t}$ production ($m_{t\bar{t}}\gtrsim1~\mathrm{TeV}$). These break the
    octet--singlet four-fermion degeneracies and strengthen constraints on
    energy-growing $qqtt$ operators.

    \item $t\bar{t}H$ and $tH$ production. These provide the most direct probe of
    $\mathcal{O}_{\phi\phi}$ and therefore directly improve its constraint.

    \item Differential single-top observables ($p_T$, rapidity), together with
    the $W$-helicity fraction $F_R$ and top polarization. These constrain the
    charged-current operators $\mathcal{O}_{tW}$, $\mathcal{O}_{bW}$, and
    $\mathcal{O}_{\phi q}^{3}$, with $F_R$ providing the cleanest handle on
    $\mathcal{O}_{bW}$.

    \item Four-top ($t\bar{t}t\bar{t}$) production. This primarily probes new
    four-heavy operators outside the current basis and provides a secondary
    cross-check of the top Yukawa sector.
\end{itemize}

Lower priority are $t\bar{t}$ spin correlations, differential $t\bar{t}W$, and
$t\bar{t}b\bar{b}$ production, which provide comparatively limited additional
constraining power.
\end{promptblock}
We then ask for a specific $t\bar{t}Z$ analysis to implement. The \sfa suggest a relatively recent ATLAS $t\bar{t}Z$ measurement~\cite{ATLAS:2023eld}. It includes inclusive and differential measurements for various observables at particle level and at parton level. Since the current global analysis only includes total rate measurements for $t\bar{t}Z$ production, this is a natural choice to increase its sensitivity. Furthermore, ATLAS provides a likelihood which gives us access to the relevant nuisance parameters, which we will discuss and use in Section~\ref{sec:likelihoods}.

\subsection{SM re-simulation}

The second re-casting step is a precise re-simulation of the SM signal. For this analysis we would only need to simulate parton-level data, but we nevertheless ask the \sfa to produce parton-level and particle-level data to show the capabilities of the agents: 
\begin{userblock}[User]
I want to implement a new measurement into the global top analysis. I've decided that I want to include the $t\bar{t}Z$ measurement from 2312.04450. I want you to begin by simulating the SM signal at both parton- and particle-level. As a final output I want plots comparing the measured data to the simulation including uncertainties, as well as all madgraph cards and python scripts needed to reproduce these results.
\end{userblock}
The agents analyze the ATLAS paper autonomously, only asking for input on choices that have a significant effect on the analysis. The only major decision that requires an expert input is the choice of kinematic observable from a list:
\begin{promptblock}[\sfa]
Which of the 17 differential observables should I simulate and plot?
\begin{itemize}
    \item[$\square$] all 17 observables — Adds the per-channel trilepton/tetralepton observables. Most complete for the global analysis; largest extraction + analysis + plotting effort.
    \item[$\square$] the $4\ell + 3\ell$ block — $p_T(Z), |y(Z)|, cos\theta^*_Z, p_T(t), p_T(t\bar{t}), |\Delta\phi(t\bar{t},Z)|, m_{t\bar{t}Z}, m_{t\bar{t}}, |y(t\bar{t}Z)|$. The EFT-sensitive, shared-fiducial physics observables most relevant to a global top analysis. Recommended balance of coverage vs effort.
    \item[$\checksquare$] A single flagship observable ($m_{\bar{t}tZ}$+$p_{T}(Z)$) —Establish the full parton+particle pipeline end-to-end on one or two flagship spectra first, validate, then expand to the rest. Fastest path to a working comparison.
\end{itemize}
Which normalization should the comparison plots use?
\begin{itemize}
    \item[$\square$] Absolute $d\sigma/dx$ — Direct data-vs-simulation comparison in physical units; includes the overall normalization (sensitive to total rate). The natural default for a data/MC comparison.
    \item[$\checksquare$] Normalized $1/\sigma d\sigma/dx$ — Shape-only comparison; cancels luminosity and much of the normalization uncertainty. Better for isolating shape modeling.
    \item[$\square$] Both absolute and normalized
\end{itemize}
\end{promptblock}
For illustration purposes we only choose $m_{t\bar{t}Z}$ and the transverse momentum $p_T(Z)$. Finally, we need to decide whether to implement the absolute or normalized measurement. We choose the normalized rate such that we can also keep the more precise total rate measurement without double counting. The \mg settings our agentic system extracts are:
\begin{itemize}
\item \textbf{Generator \quad} MadGraph5\_aMC@NLO at NLO;
\item \textbf{PDF set \quad} NNPDF3.0nlo (LHAPDF ID 260000);
\item \textbf{Scales \quad} Dynamic, $\mu_R=\mu_F=H_T/2$ with $H_T = \sum_i\sqrt{(m_i^2+p_{T,i}^2)}$;
\item \textbf{Decays \quad} $\tau\rightarrow\ell$ excluded (direct $W/Z\rightarrow e/\mu$ only).
\end{itemize}
The particle-level fiducial cuts extracted by our agent are shown in Table~\ref{tab:sel_cuts}.

\begin{table}[b!]
\centering
\label{tab:particle_selection}
\renewcommand{\arraystretch}{1.25}
\begin{small} \begin{tabular}{@{}lcc@{}}
\toprule
\multicolumn{3}{c}{\textbf{Common selection cuts
}} \\
\midrule
Objects &
\multicolumn{2}{c}{$e,\mu$: dressed ($\Delta R=0.1$), $|\eta|<2.5$;\quad
jets: anti-$k_t$ ($R=0.4$), $|\eta|<2.5$} \\
Jets &
\multicolumn{2}{c}{$p_T>25$ GeV,\quad
$\ge1$ ghost-matched $b$-jet ($p_T>5$ GeV)} \\
$Z$ candidate &
\multicolumn{2}{c}{$\ge1$ OSSF pair with $|m_{\ell\ell}-m_Z|<10$ GeV} \\
\midrule
Leptons & \textbf{$3\ell$} & \textbf{$4\ell$} \\
\midrule
Leading $p_T$ [GeV]
& $27/20/15$
& $27/7/7/7$ \\
Charge
& $|\sum q|=1$
& $\sum q=0$ \\
Jets
& $\ge3$
& $\ge2$ \\
Additional
& --
& All OSSF pairs satisfy $m_{\ell\ell}>10$ GeV \\
\bottomrule
\end{tabular} \end{small}
\caption{Particle-level event selection for the 3-lepton and 4-lepton signal regions.}
\label{tab:sel_cuts}
\end{table}

\begin{figure}[t]
    \includegraphics[width=1\linewidth]{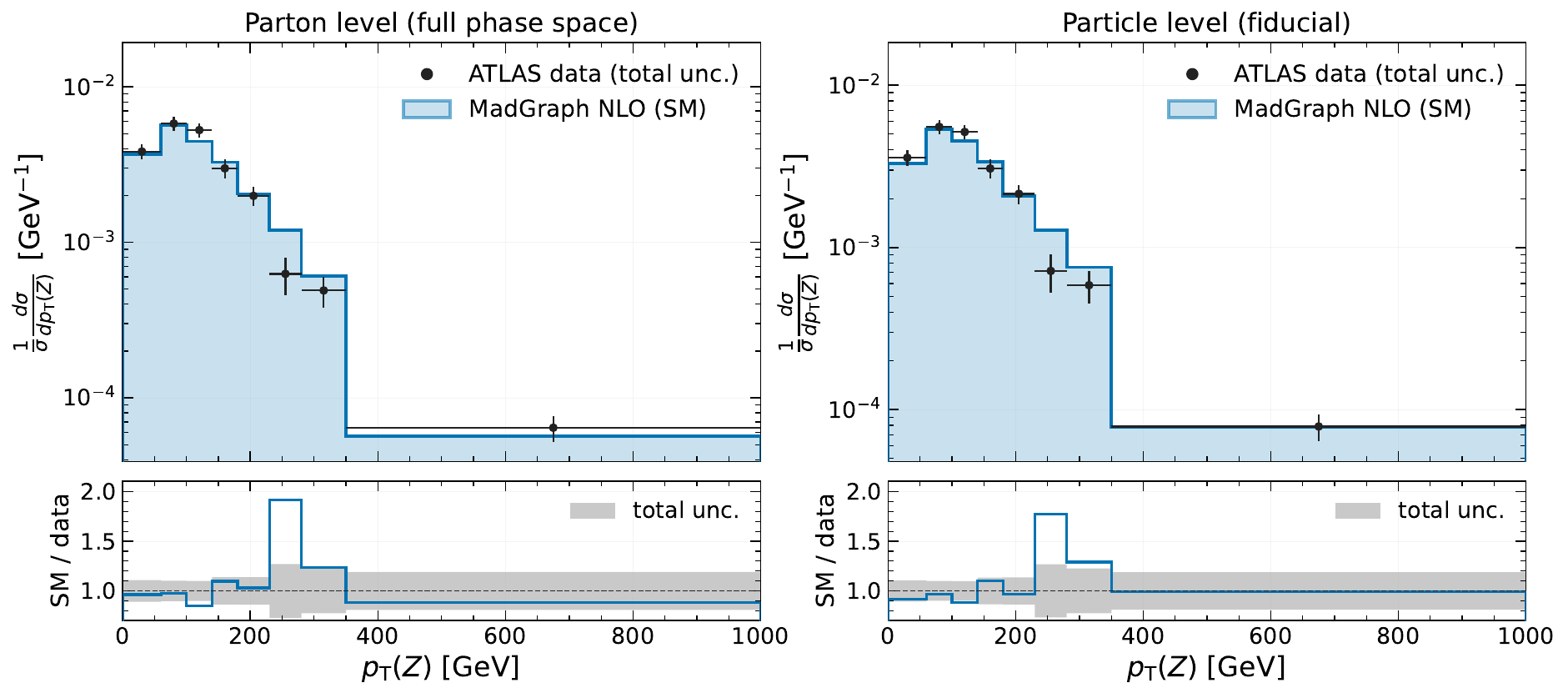} \\
    \includegraphics[width=1\linewidth]{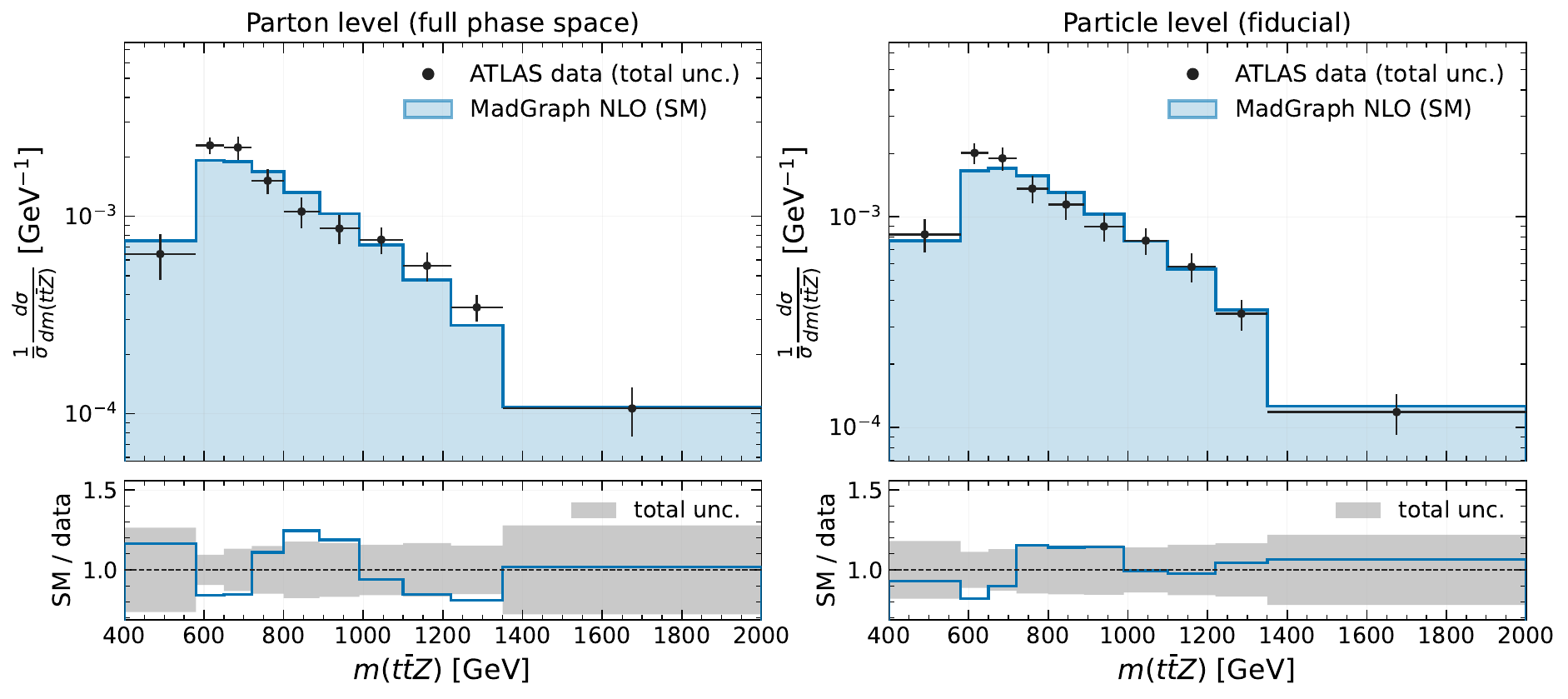}
    \caption{Comparison of simulation vs data at parton level (left) and particle level (right) for $p_T(Z)$ (top) and the  $m_{t\bar{t}Z}$ (bottom).}
    \label{fig:SM_data_comp}
\end{figure}

In the top panels of Figure~\ref{fig:SM_data_comp} we compare the simulated signal to the ATLAS measurement at parton level and at particle level. We find very good agreement in all bins but one, where the expected value is larger by approximately $3\sigma$. In the bottom panels we show the same for $m_{t\bar{t}Z}$, finding very good agreement everywhere within uncertainties. All panels perfectly reproduce their counterparts in the ATLAS publication.

\subsection{SMEFT simulation}

Next, we ask the \sfa to add SMEFT operators for our new signal hypothesis:
\begin{userblock}[User]
For the next step, simulate the SMEFT effects needed for the parametrization of their effect in out global analysis. As the final output I want plots similar to those comparing parton- and particle-level simulation to measured data as before, this time displaying the effect of the different Wilson coefficients on these observables. Naturally, I also want you to use these results to create the \sfitter JSON file for this measurement.
\end{userblock}
Once again, the \sfa ask for decisions with significant impact on the analysis, now about the complexity and the number of dimension-6 operators. The agents give three options:
\begin{promptblock}[\sfa]
Which Wilson coefficients should I parametrize for $t\bar{t}Z$?
\begin{itemize}
    \item[$\square$] $c_{tZ}, c_{\phi t}, c_{\phi Q}^{-}, c_{tW}$ (top-Z/W couplings ttZ directly constrains) + $c_{tG}$ (chromomagnetic) — The natural set for the WC-effect plots and a focused  datacard.
    \item[$\checksquare$] The full global $t\bar{t}Z$ set (21 operators) — Complete for the global analysis, but most are weakly constrained by $t\bar{t}Z$ and it's a much larger scan campaign.
    \item[$\square$]  Dipoles only ($c_{tZ}, c_{tW}$) Just the top electroweak dipoles — the steepest tail-shaping operators. Minimal campaign, fastest, 
     but a narrow constraint.
\end{itemize}
\end{promptblock}
We choose the full set to challenge our agentic setup. The \sfa follow up on the precision of the SMEFT simulation:
\begin{promptblock}[\sfa]
What EFT order / particle-level scope? (This is the main compute driver — particle-level NLO EFT is the expensive part.)
\begin{itemize}
    \item[$\square$] NLO at both levels — SMEFT at NLO QCD (consistent with the NLO SM baselines), parton + particle. Full and rigorous; ~11 fixed-order runs (cheap) + ~11 NLO+PS samples (parallelized on the cluster, a few hours wall). Recommended.
    \item[$\checksquare$] NLO parton + reuse $\kappa$ for particle plots — Full NLO EFT at parton level; for the particle-level plots, reuse the parton-level $\kappa$ shape (assumes the fiducial acceptance doesn't strongly change the WC shape response — a common, flagged approximation). Much faster.
    \item[$\square$]  Scan the EFT at LO (SM baseline stays NLO) — Fastest for both levels; a widely-used approximation but less consistent with the NLO baseline.
\end{itemize}
\end{promptblock}
In the following section we will use the parton-level data and simulations to constrain the Wilson coefficients. This is why we ask the \sfa to only simulate parton-level at NLO and reuse their $\kappa$ for the particle plots.

\begin{figure}[b!]
    \includegraphics[width=1\linewidth]{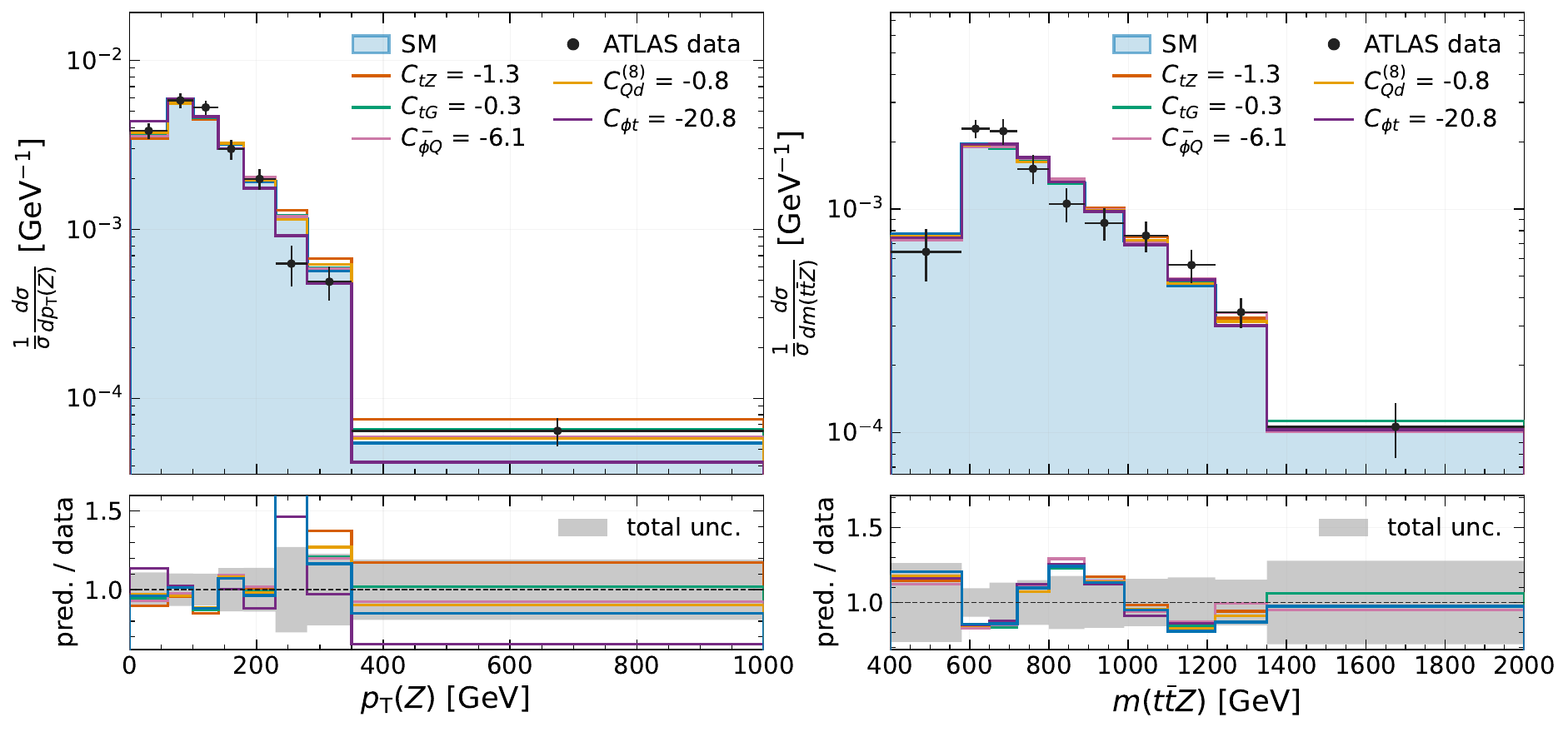}
    \caption{Effect of a selection of Wilson coefficients chosen to deviate by 3$\sigma$.}
    \label{fig:SMEFT_obs}
\end{figure}

The results of the SMEFT simulations are shown in Figure~\ref{fig:SMEFT_obs} for a selection of Wilson coefficients. We see a particularly strong effect in the tails of the kinematic distributions, as expected from the energy or momentum dependence. We also see that the deviations are much larger for $p_{T}(Z)$ than for $m_{t\bar{t}Z}$ given the same Wilson coefficients.

A particularly interesting outcome is that to match the significantly lower single bin we need a large negative value of $C_{\phi t}$. This anomalous bin will correspondingly pull the global analysis away from the SM.

\subsection{Global analysis}

The last \sfitter step is the statistical analysis of the exclusive likelihood over the Wilson coefficients:
\begin{userblock}[User]
Finally, I want you to use the new datacard to run an analysis for just the new measurement alone first, after that I want you to run the full global analysis once with and once without the new measurement. As a final output I want a plot showing the constraints for the individual measurement, as well as a plot comparing the constraints from the global analysis with and without the new measurement.
\end{userblock}
\begin{promptblock}[\sfa]
Both fits are done and both plots are delivered.\\

What this shows — $p_T(Z)$ does essentially all the work. ~2.7× tighter than $m(ttZ)$ on $C_{tZ}$ standalone, and in the global analysis it delivers the entire impact: 34–42\% on the three top-EW operators, while $m(ttZ)$ gives 19\% on $C_{tZ}$ and nothing elsewhere. The sub-2\% wiggles on the other 19 are noise, not signal - several are negative, and adding data cannot loosen a bound.
\end{promptblock}
In Figure~\ref{fig:SMEFT_obs} we see that the four-fermion operators, like $O_{Qd}^{(8)}$, are sensitive to the $t\bar{t}Z$ kinematics through their energy dependence, but are already strongly constrained through the $t\bar{t}$ measurements. This is especially true once we include boosted top kinematics. This is why we do not expect these operators to get constrained more strongly. Similarly, $O_{tG}$ is constrained by total rate measurements and less sensitive to the normalized kinematic distributions.

\begin{figure}[t]
    \includegraphics[width=1\linewidth]{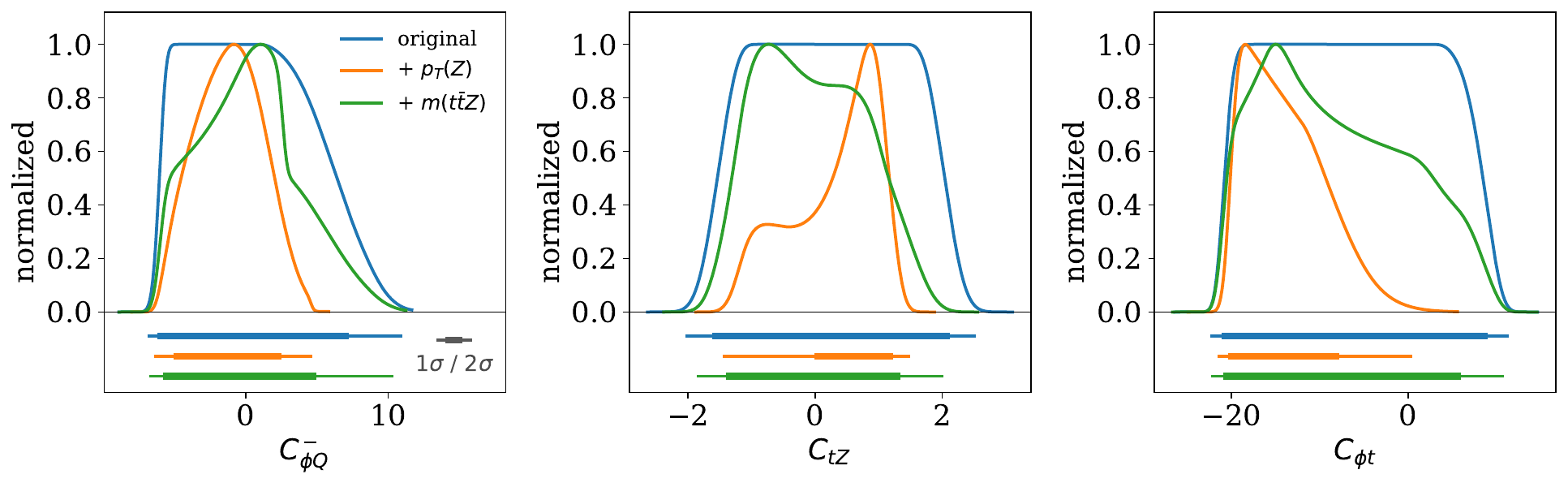}
    \caption{Profiled constraints on Wilson coefficients from the original dataset and after adding the $p_{T}(Z)$ or $m_{t\bar{t}Z}$ kinematic information.}
    \label{fig:marginals_new_ttZ}
\end{figure}

\begin{figure}[b!]
    \includegraphics[width=1\linewidth]{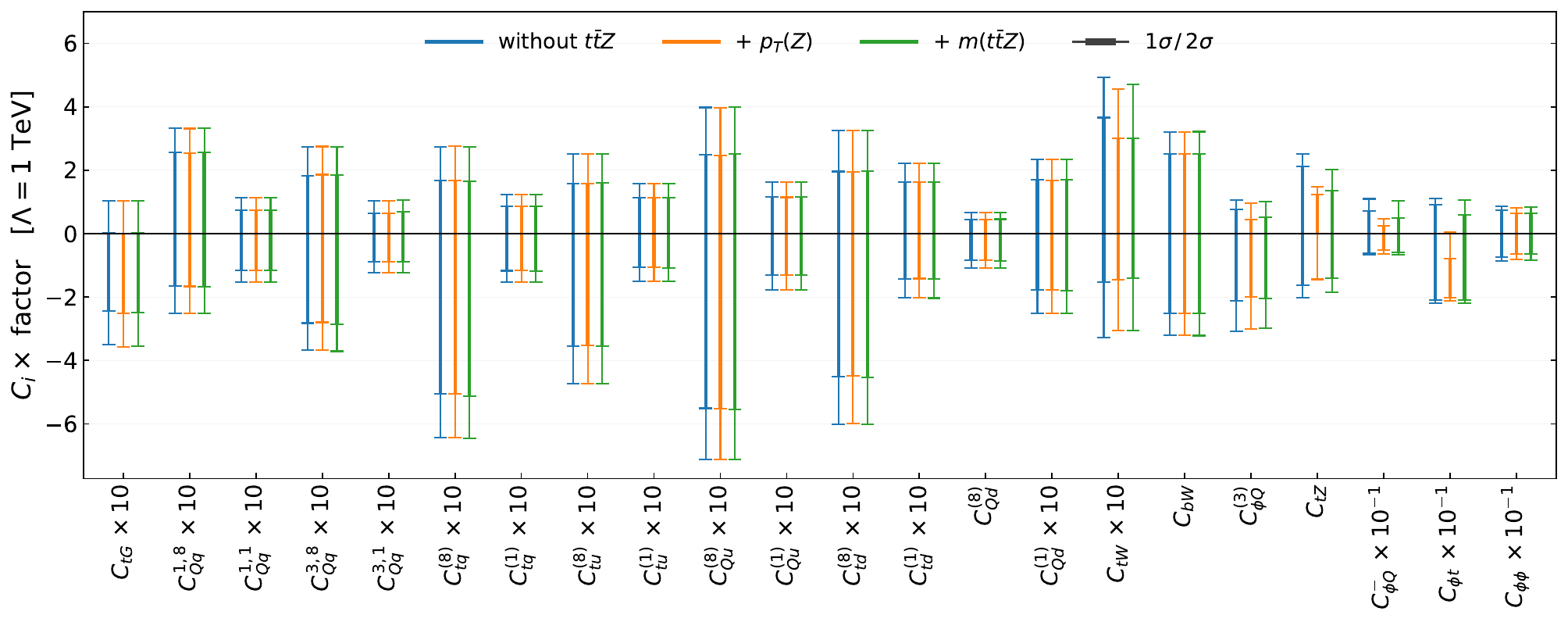}
    \caption{Profiled constraints before and after adding the $p_{T}(Z)$ or $m_{t\bar{t}Z}$ kinematic information.}
    \label{fig:marg_confidence_new}
\end{figure}

In Figure~\ref{fig:marginals_new_ttZ} we show how the constraints tighten when we add the $p_{T}(Z)$ or $m_{t\bar{t}Z}$ measurements to the original analysis. We choose three operators that are constrained by $t\bar{t}Z$ production at different levels, but not by the much more constraining $t\bar{t}$ process. Without $t\bar{t}Z$ kinematics, $C_{\phi t}$ has the most remaining flexibility especially towards large negative Wilson coefficients. Strong constraints arise when we include either of the two observables, although those from $p_T(Z)$ are significantly stronger. Looking more closely, the central value for $C_{\phi t}$ is pulled far away from the SM. This is the result of the $3\sigma$ underfluctuation, pushing $C_{\phi t}$ to large negative values. This effect is absent for the $m_{t\bar{t}Z}$ measurement where no underfluctuation is observed.

Finally, we show the effect of the new kinematic $t\bar{t}Z$ measurements on the global \sfitter analysis in Figure~\ref{fig:marg_confidence_new}. All individual results are profiled. In line with the above discussion, the new measurement has no visible effect on any operators constrained by the $t\bar{t}$ total rate and boosted kinematics. This figure would be the final result of an \sfitter top-sector SMEFT analysis with the new $t\bar{t}Z$ kinematic measurements included.

\subsection{Published likelihoods}
\label{sec:likelihoods}

\begin{figure}[b!]
    \includegraphics[width=1\linewidth]{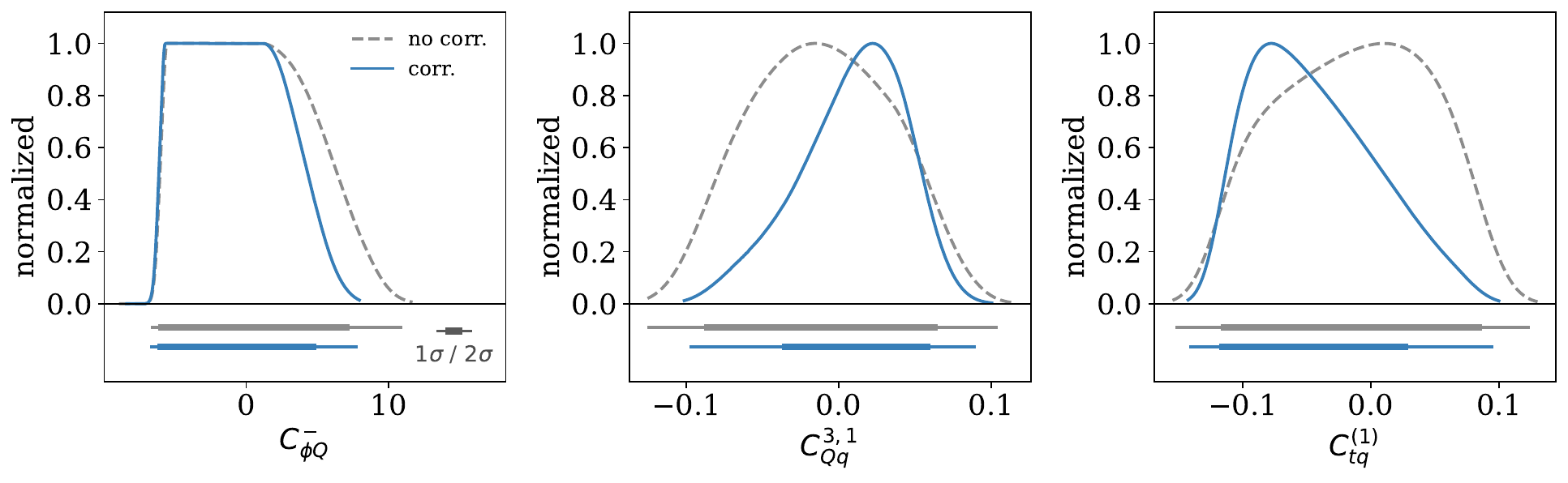}
    \caption{Effect of correlations between systematics on the profiled global analysis constraints.}
    \label{fig:corr_effects}
\end{figure}

One strength of \sfitter is that it takes into account the correlation of systematic uncertainties between experiments. To emphasize their importance, we show their effect on select coefficients in our global analysis in Figure~\ref{fig:corr_effects}, leading to non-negligible shifts in the constraints. In the previous sections we neglected the detailed discussion of their implementation.

There are different ways to extract and implement uncertainties for a measurement: First, we can extract the total uncertainty for each bin from the HEPData entry. This is accurate in the per-bin split of the uncertainty, but it combines systematic and statistical uncertainties and makes it impossible to correlate systematics between measurements. Second, we can take the uncertainty split provided in Table~14 of Ref.~\cite{ATLAS:2023eld}. This gives us the split of systematic uncertainties, but only for the total rate, so we need to somehow estimate the per-bin uncertainties. This is close to the \sfitter approach we take when no more information is provided. Third, we can use the bin-to-bin covariance matrix published by the experiment. This gives us more information, but it again conflates statistical and systematic uncertainties and makes it impossible to correlate the systematics the way we want to do it in \sfitter and can lead to negative correlations incompatible with the fully correlated systematics currently implemented.

\begin{figure}[t]
    \includegraphics[width=1\linewidth]{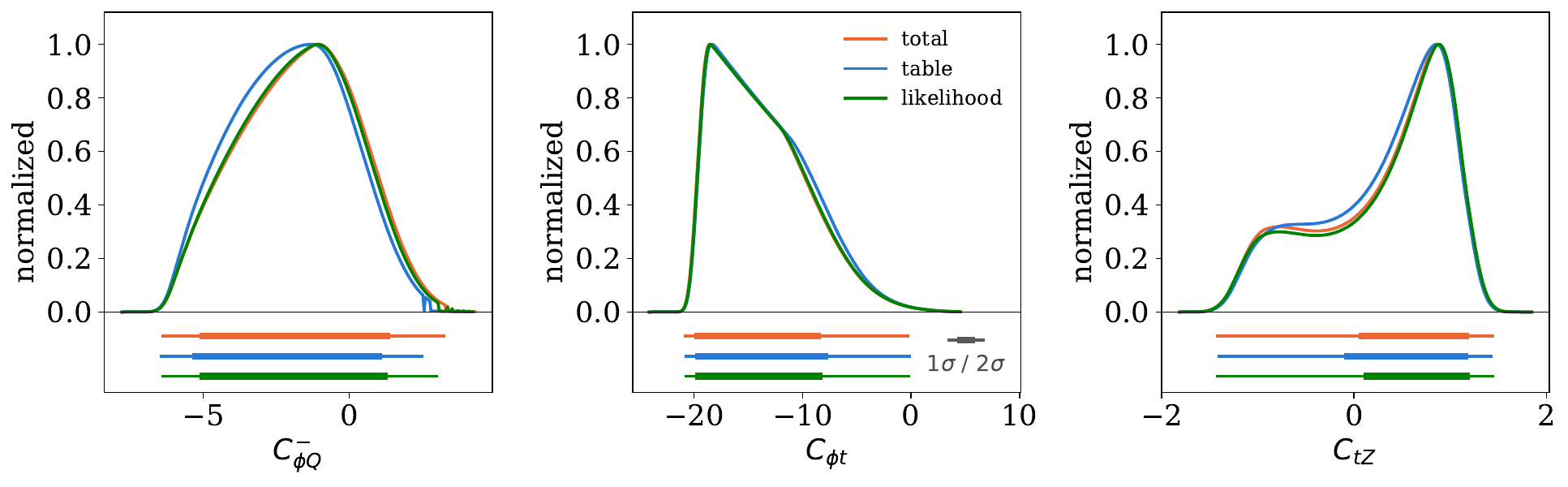}
    \caption{Difference between implementing a single total uncertainty, estimating per-bin uncertainties from the table or using the public likelihood.}
    \label{fig:corr_methods}
\end{figure}

In contrast to all these ad-hoc solutions, public likelihoods provide us with much greater control over uncertainties. We can now take into account all correlations between nuisance parameters, 276 systematic nuisance parameters for the $t\bar{t}Z$ measurement. We can now  compute the uncertainties for each bin by grouping the nuisance parameters following our \sfitter convention and performing a profiled likelihood fit. We show the result of a global analysis using $(i)$ the total uncertainty, $(ii)$ the per-bin estimated uncertainties from the table, and $(iii)$ the uncertainties extracted from a profile likelihood fit for three critical Wilson coefficients in Figure~\ref{fig:corr_methods}. There is barely any visible difference, reflecting the statistics-dominated uncertainty budget of the normalized kinematic distributions.

While the use of a public likelihood does not change the final analysis result dramatically for the $t\bar{t}Z$ kinematic measurement, this will be different for systematics-dominated analyses. The \sfa can handle the hundreds of nuisances provided by the likelihoods reliably. 

\section{Outlook and Impact}
\label{sec:conclusion}

LHC simulations, re-simulations, and analysis re-casting cannot be automized and run by non-experts (yet), but they involve many steps which benefit from agentic support of users at all levels of experience. In this way they provide a perfect benchmark for the accelerating impact of interactive support agents in a realistic LHC setting.

We first presented \mav{3} as a new implementation with a focus on trustworthy support at all levels of user experience. Unlike the earlier versions~\cite{Plehn:2026gxv}, it is based on the more economical consultant architecture. 

Building on these simulation agents, we introduced the new \sfa to support analysis re-casting. For a global SMEFT analysis of the top sector we showed how the \sfa can choose, implement, and evaluate an additional $t\bar{t}Z$ kinematic measurement. Each step in the re-casting workflow is supervised by a human physicist. This proof of concept will eventually allow us to expand the set of analyses implemented in \sfitter efficiently and with no risk concerning the quality of the results.

The effectiveness of the \sfa raises exciting questions about useful ways to publish experimental results. Obviously, if high-dimensional correlations are not reported, they are lost, motivating public unfolded data. However, documenting all available information in any form, for instance, thousands of pages of hardly structured documentation combined with source codes and data cards, should be sufficient for LLM agents to re-cast an analysis. What remains crucial is that \sfa provide ways for human physicists to check and learn from the agentic work step by step.

\section*{Acknowledgments}

We would like to thank Dario H\"ugel, Erik Gerwick, and the barra team for their inspiring advice. We would also like to thank Siddharth Mishra-Sharma for providing us with Anthropic Claude access and for his extremely helpful advice on our agentic path. This research is supported by the Deutsche Forschungsgemeinschaft (DFG, German Research Foundation) under grant 396021762--TRR~257: \textsl{Particle Physics Phenomenology after the Higgs Discovery}. We acknowledge support by the state of Baden-Württemberg through bwHPC and the German Research Foundation (DFG) through grant no INST 39/963-1 FUGG (bwForCluster NEMO). DS is funded by Germany's Excellence Strategy EXC~2181/1 -- 390900948 (the \textsl{Heidelberg STRUCTURES Excellence  Cluster}). This project and SD were supported by the Baden-Württemberg Stiftung.

\appendix
\section{Top sector measurements}
\label{app:top_data}

\begin{table}[h!]
\centering
\begin{small} \begin{tabular}{lcccccc}
        \toprule
        Experiment & $\sqrt{s}$ [TeV] & $\mathcal{L}$ [fb$^{-1}$] & Channel & Observable & \# Bins & Ref.\\
        \midrule
        CMS & 8 & 19.7 & $e \mu$ & $\sigma_{t \bar{t}}$ & & \cite{CMS:2016yys}\\
        ATLAS & 8 & 20.2 & $lj$ & $\sigma_{t \bar{t}}$ & & \cite{ATLAS:2017wvi}\\
        \midrule
        CMS & 13 & 137 & $lj$ & $\sigma_{t \bar{t}}$ & & \cite{CMS:2021vhb}\\
        CMS & 13 & 35.9 & $ll$ & $\sigma_{t \bar{t}}$ & & \cite{CMS:2018fks}\\
        ATLAS & 13 & 36.1& $ll$ & $\sigma_{t \bar{t}}$ & & \cite{ATLAS:2019hau}\\
        ATLAS & 13 & 36.1 & $aj$ & $\sigma_{t \bar{t}}$ & & \cite{ATLAS:2020ccu}\\
        ATLAS & 13 & 139 & $lj$ & $\sigma_{t \bar{t}}$ & & \cite{ATLAS:2020aln}\\
        \midrule
        CMS & 13.6 & 1.21 & $ll$, $lj$ & $\sigma_{t \bar{t}}$ & & \cite{CMS:2023qyl}\\
        \midrule
        CMS & 8 & 19.7 & $lj$ & $\frac{1}{\sigma} \frac{d \sigma}{d p_{T}^{t}}$ & 7 & \cite{CMS:2015rld}\\
        CMS & 8 & 19.7 & $l l$ & $\frac{1}{\sigma} \frac{d \sigma}{d p_{T}^{t}}$ & 5 & \cite{CMS:2015rld}\\
        ATLAS & 8 & 20.3 & $lj$ & $\frac{1}{\sigma} \frac{d \sigma}{d m_{t \bar{t}}}$ & 7 & \cite{ATLAS:2015lsn}\\
        \midrule
        CMS & 13 & 137 & $lj$ & $\frac{1}{\sigma} \frac{d \sigma}{d m_{t \bar{t}}}$ & 15 & \cite{CMS:2021vhb}\\
        CMS & 13 & 35.9 & $ll$ & $\frac{1}{\sigma} \frac{d \sigma}{d \Delta y_{t \bar{t}}}$ & 8 & \cite{CMS:2018adi}\\
        ATLAS & 13 & 36 & $lj$ & $\frac{1}{\sigma} \frac{d \sigma}{d m_{t \bar{t}}}$ & 9 & \cite{ATLAS:2019hxz}\\
        ATLAS & 13 & 139 & $aj$, high-$p_{T}$& $\frac{1}{\sigma} \frac{d \sigma}{d m_{t \bar{t}}}$ & 13 & \cite{ATLAS:2022mlu}\\
        \midrule
        CMS & 8 & 19.7 & $lj$ & $A_{C}$ & & \cite{CMS:2015pob}\\
        CMS & 8 & 19.5 & $ll$ & $A_{C}$ & & \cite{CMS:2016ypc}\\
        ATLAS & 8 & 20.3 & $lj$ & $A_{C}$ & & \cite{ATLAS:2015jgj}\\
        ATLAS & 8 & 20.3 & $ll$ & $A_{C}$ & & \cite{ATLAS:2016ykb}\\
        \midrule
	    CMS & 13 & 138 & $lj$ & $A_{C}$ & & \cite{CMS:2022ged}\\
	    ATLAS & 13 & 139 & $lj$ & $A_{C}$ & & \cite{ATLAS:2022waa}\\
        \midrule
        ATLAS & 13 & 139 & & $\sigma_{t \bar{t} Z}$ & & \cite{ATLAS:2021fzm}\\
        CMS & 13 & 77.5 & & $\sigma_{t \bar{t} Z}$  & & \cite{CMS:2019too}\\
        \midrule 
        ATLAS & 13 & 140 & $3\ell + 4\ell$ & $\frac{1}{\sigma} \frac{d \sigma}{dp_{T}(Z)}$ & 8 & \cite{ATLAS:2023eld}\\
        ATLAS & 13 & 140 & $3\ell + 4\ell$ & $\frac{1}{\sigma} \frac{d \sigma}{dm_{t\bar{t}Z}}$ & 8 & \cite{ATLAS:2023eld}\\
	    \midrule
	    CMS & 13 & 35.9 & & $\sigma_{t \bar{t} W}$ & & \cite{CMS:2017ugv}\\
	    ATLAS & 13 & 36.1 & & $\sigma_{t \bar{t} W}$ & & \cite{ATLAS:2019fwo}\\
        \midrule
	    CMS & 8 & 19.7 & & $\sigma_{t \bar{t} \gamma}$ & & \cite{CMS:2017tzb}\\
        ATLAS & 8 & 20.2 & & $\sigma_{t \bar{t} \gamma}$ & & \cite{ATLAS:2017yax}\\
        \bottomrule
\end{tabular} \end{small}
\caption{Top pair observables included in our global analysis, adapted from the previous \sfitter
  analysis~\cite{Brivio:2019ius}.}
\label{tab:datasets_ttbar}
\end{table}

\begin{table}[b!]
\centering
	\begin{small} \begin{tabular}{lcccccc}
        \toprule
		Exp. & $\sqrt{s}$ [TeV] & $\mathcal{L}$ [fb$^{-1}$] & Channel & Observable & \# Bins & Ref.\\
        \midrule
		ATLAS & 7 & 4.59 &$t$-ch & $\sigma_{tq + \bar{t} q}$  & & \cite{ATLAS:2014sxe}\\
    CMS & 7 & \footnotesize{1.17 ($e$), 1.56 ($\mu$)} &$t$-ch & $\sigma_{tq + \bar{t} q}$  & & \cite{CMS:2012xhh}\\
		ATLAS & 8 & 20.2 &$t$-ch & $\sigma_{tq}$, $\sigma_{\bar{t}q}$ & & \cite{ATLAS:2017rso}\\
		CMS & 8 & 19.7 &$t$-ch & $\sigma_{tq}$, $\sigma_{\bar{t}q}$  & & \cite{CMS:2014mgj}\\
		ATLAS & 13 & 3.2 &$t$-ch & $\sigma_{tq}$, $\sigma_{\bar{t}q}$ & & \cite{ATLAS:2016qhd}\\
		CMS & 13 & 2.2 &$t$-ch & $\sigma_{tq}$, $\sigma_{\bar{t}q}$ & & \cite{CMS:2016lel}\\
		CMS & 13 & 35.9 &$t$-ch & $\frac{1}{\sigma} \frac{d \sigma}{d |p_{T,t}|}$  & 5 & \cite{CMS:2019jjp} \\
	\midrule 
		CMS & 7 & 5.1 & $s$-ch & $\sigma_{t \bar{b} + \bar{t} b}$ & & \cite{CMS:2016xoq}\\
		CMS & 8 & 19.7 & $s$-ch & $\sigma_{t \bar{b} + \bar{t} b}$ & & \cite{CMS:2016xoq} \\
		ATLAS & 8 & 20.3 & $s$-ch & $\sigma_{t \bar{b} + \bar{t} b}$ & & \cite{ATLAS:2015jmq} \\
		ATLAS & 13 & 139 & $s$-ch & $\sigma_{t \bar{b} + \bar{t} b}$ & & \cite{ATLAS:2022wfk} \\
	\midrule
		ATLAS & 7 & 2.05 & $tW$ $(2l)$ & $\sigma_{t W + \bar{t} W}$ & & \cite{ATLAS:2012bqt}\\
		CMS & 7 & 4.9 & $tW$ $(2l)$ & $\sigma_{t W + \bar{t} W}$ & & \cite{CMS:2012pxd} \\
		ATLAS & 8 & 20.3 & $tW$ $(2l)$ & $\sigma_{t W + \bar{t} W}$ & & \cite{ATLAS:2015igu} \\
		ATLAS & 8 & 20.2 & $tW$ $(1l)$ & $\sigma_{t W + \bar{t} W}$ & & \cite{ATLAS:2020cwj}\\
		CMS & 8 & 12.2 & $tW$ $(2l)$ & $\sigma_{t W + \bar{t} W}$ & & \cite{CMS:2014fut}\\
		ATLAS & 13 & 3.2 & $tW$ $(1l)$ & $\sigma_{t W + \bar{t} W}$ & & \cite{ATLAS:2016ofl}\\
		CMS & 13 & 35.9 & $tW$ $(e \mu j)$  & $\sigma_{t W + \bar{t} W}$ & & \cite{CMS:2018amb} \\
		CMS & 13 & 36 & $tW$ ($2l$) & $\sigma_{t W + \bar{t} W}$ & & \cite{CMS:2021vqm}\\
    \midrule
        ATLAS & 13 & 36.1 & $tZ$ & $\sigma_{t Zq}$ & & \cite{ATLAS:2017dsm}\\
	\midrule
        ATLAS & 7 & 1.04 &  & $F_0$,$F_L$ & & \cite{ATLAS:2012nhi}\\
	    CMS & 7 & 5 & & $F_0$,$F_L$ & & \cite{CMS:2013xxb}\\
        ATLAS & 8 & 20.2 &  & $F_0$,$F_L$ & & \cite{ATLAS:2016fbc}\\
	    CMS & 8 & 19.8 &  & $F_0$,$F_L$ & & \cite{CMS:2016asd}\\
	    ATLAS & 13 & 139 & & $F_0$,$F_L$ & & \cite{ATLAS:2022rms}\\
        \bottomrule
\end{tabular} \end{small}
\caption{Single top and top decay observables included in our global
  analysis, adapted from the previous \sfitter
  analysis~\cite{Brivio:2019ius}.}
\label{tab:datasets_singletop}
\end{table}

\clearpage
\section{Silent failure tests}
\label{app:silent-failure-benchmark}

A silent failure refers to a \mg configuration or execution in which no error messages are produced while the simulated physics differs from what was requested. The updated \mav{3} is built to prevent them. We formulate five questions that may lead to silent failures. The questions are phrased from the perspective of a physicist. We deliberately design the questions so that they are not directly addressed in tutorials or other publicly available sources and instead require reasoning:
\begin{enumerate}
\item \textbf{Conjugate decays \quad} First, we ask for the final state particles of a  decay chain:

\begin{userblock}[User]
I'm running \texttt{generate p p > t t\textasciitilde{}, t > b w+} in MG5 (default \texttt{sm}, default cards). What are the matrix-element-level final-state particles of each event?
\end{userblock}

The correct answer is $\{b, W^+, \bar t\}$, as \mg applies the decay chain to all identical particles, but not to the charge conjugated ones.

\item \textbf{Decay nesting: \quad} Next, we ask for a fully nested decay-chain command, in which the same particle appears at two different depths:

\begin{userblock}[User]
I want to set up the following process as an explicit decay chain with the SM model:
$$
pp \to W^+ H, \quad H \to t\bar t, \quad t \to b\,W^+, \quad \bar t \to \bar b\,W^-.
$$
The $W^+$ produced in association with the Higgs decays leptonically ($W^+ \to \ell^+\nu$), the $W^+$ from the top decays hadronically ($W^+ \to jj$), and the $W^-$ from the antitop decays leptonically ($W^- \to \ell^-\bar\nu$), giving an $\ell^+\nu\,\ell^-\bar\nu\,b\bar b\,jj$ final state.

What is the full decay-chain process command for this? Every decay listed above must appear in the command itself; do not defer any of them to MadSpin or the parton shower.
\end{userblock}

We note that for Standard Model masses, this decay is highly off-shell. However, since we do not specify them, this setup could be run with a larger Higgs mass. The correct process command involves multiple nested chains:
\begin{center}
\texttt{generate p p > w+ h, w+ > l+ vl, (h > t t\textasciitilde{}, (t > b w+, w+ > j j), (t\textasciitilde{} > b\textasciitilde{} w-, w- > l- vl\textasciitilde{}))}
\end{center}
\mg matches each decay clause to the identical particles within its parenthesised group, so the two $W^+$ are distinguished only by nesting. Dropping the parentheses would result in a different decay chain.

\item \textbf{Fiducial cuts \quad} We generate high-mass-tail events for $W^+ \to \ell^+\nu$ in two apparently equivalent ways and observe that they lead to significantly different cross sections, so we ask for a diagnosis:

\begin{userblock}[User]
I'm computing the high-mass tail of $W^+ \to \ell^+\nu$ at the LHC ($\sqrt{s} = 13$ TeV) --- the lepton--neutrino invariant-mass region $M(\ell\nu) > 200$ GeV, well above the $W$ pole --- with the run-card's default lepton cuts (\texttt{ptl}, \texttt{etal}) in place. I wrote the process two ways and the results don't make sense to me.

The s-channel form \texttt{generate p p > w+ > l+ vl} gives $\sigma \approx 6.79$ pb, and I get the same number whatever I set \texttt{bwcutoff} to.

The decay-chain form \texttt{generate p p > w+, w+ > l+ vl} behaves completely differently. At the default \texttt{bwcutoff = 15} it returns \textbf{zero} cross-section. As I raise \texttt{bwcutoff} the cross-section climbs --- about $5.5$ pb at \texttt{bwcutoff = 100}, $7.6$ pb at 300 --- and then it plateaus at \textbf{7.67 pb} for \texttt{bwcutoff} of 1000 or more.

So at large \texttt{bwcutoff} the decay-chain form settles at $7.67$ pb --- but that still isn't the $6.79$ pb the s-channel form gives. I expected that opening \texttt{bwcutoff} wide enough would make the two forms agree, and it doesn't.

Why does the decay-chain form depend on \texttt{bwcutoff} while the s-channel form doesn't, and why --- even at large \texttt{bwcutoff} --- is there still a residual gap between $7.67$ and $6.79$? Which of the two numbers is the physical fiducial cross-section I want?
\end{userblock}

By default, \mg does not apply fiducial cuts to decay products, which has to be manually enabled. The corresponding logic is buried deep in the source code.

\item \textbf{Off-shell decay \quad} We ask for the LO setup of a BSM heavy-pseudoscalar signal:

\begin{userblock}[User]
I'm setting up the LO signal calculation for a 2HDM type-II heavy-pseudoscalar search at the LHC ($\sqrt{s} = 13$ TeV). The signal is $$b\bar b \to A, \qquad A \to Z\,h, \qquad Z \to \ell^+\ell^-, \qquad h \to b\bar b$$ in 2HDM type-II at large $\tan\beta$, where $A$ is the CP-odd pseudoscalar with $m_A = 170$ GeV (below the on-shell $Zh$ threshold, so we're in the off-shell regime). The SM-like Higgs is at $m_h = 125$ GeV. Parton level only. What MG5 model do I import, what process command should I feed \texttt{MG5\_aMC>}, and which \texttt{param\_card.dat} / \texttt{run\_card.dat} changes do I need? I'll do the final generation run myself --- just give me the setup with all the parameter specs I need to apply.
\end{userblock}

A natural setup for this process is to specify the decay cascade using the decay-chain syntax. In this case, \mg uses a truncated Breit--Wigner propagator for the intermediate particles, restricting their virtual invariant mass to within $15$ widths of their on-shell mass. The truncation removes part of the relevant phase space and reduces the cross section. One possible solution is to widen the truncation.

\item \textbf{SMEFT setup \quad} Finally, we ask for leptonic $t\bar t$ production with a SMEFT dipole operator using the SMEFTatNLO model:

\begin{userblock}[User]
I want the LO cross-section at $\sqrt{s} = 13$ TeV for $pp \to t\bar t \to b\,\ell^+\nu_\ell\;\bar b\,\ell^-\bar\nu_\ell$ (dileptonic, both tops decaying via the $t \to bW^+, W^+ \to \ell^+\nu_\ell$ chain), including the contribution of the SMEFT top--W dipole operator $O_{tW} = (\bar Q\sigma^{\mu\nu}\tau^I t)\tilde\phi W^I_{\mu\nu}$ at $c_{tW} = 1.0$, $\Lambda = 1$ TeV, with all other Wilson coefficients set to zero. Use the SMEFTatNLO model. What process command + \texttt{param\_card.dat} / \texttt{run\_card.dat} changes do I need? I'll do the final generation run myself --- just give me the setup with all the parameter specs I need to apply.
\end{userblock}

This question hides two challenges. First, \mg truncates the perturbative expansion already at tree level. The SMEFTatNLO model contains the single dimension-six four-fermion contact interaction $q\,q \to t\bar t$. In the default truncation, the top pair is produced only through this vertex, even if its Wilson coefficient is set to $0$, and the requested dipole operator is neglected as well. To include the Standard Model top pair production and the dipole operator in the simulation, the truncation has to be adjusted. In addition, the dipole operator significantly modifies the $t \to bW$ decay rate, consequently increasing the top width. For a consistent simulation, the top width has to be recomputed.
\end{enumerate}
For each question, we run setups that produce the silent failures and setups that do not. Comparing them, we determine which output separates them, such as a shift in the cross section, a Feynman diagram that appears or drops out, or a change in the final state. The grading of the answers is based on those references.

For validation, we use the self-hosted \href{https://huggingface.co/Qwen/Qwen3.6-27B}{Qwen3.6-27B} model, run through the Claude Code harness. We use the model without quantization and use the recommended temperature of $1.0$. Hosting the model locally ensures that all runs use the same model implementation and inference settings. The validation is then reproducible up to the run-to-run variation of temperature-one sampling, so we answer every question ten times.

An additional agent, running on Claude Opus 4.8, manages the validation procedure from start to finish. It presents each test question to the setup under evaluation, collects the response and grades it, potentially invoking \mg to that end. Only a requirement the prompt states can make an answer wrong. A choice it leaves open, such as how a particle is produced, never does.

We compare three setups:
\begin{enumerate}
\item \textbf{Bare}: unmodified Claude Code harness;
\item \textbf{Cold}: \ma with empty memories and lessons;
\item \textbf{Warm}: \ma with accumulated memories and lessons.
\end{enumerate}
The bare setup provides a baseline for the model's capabilities. The cold setup isolates the effect of the \ma architecture, while the warm setup additionally captures the effect of accumulated knowledge.
To build the memories and lessons used in the warm setup, we first put \ma through a dedicated training phase. It includes a study of the \mg source code with a focus on potentially misleading internal behavior, a set of additional questions for which \ma receive explicit feedback whenever they produce a silent failure, and a study of the curated documentation from the previous \ma version~\cite{Plehn:2026gxv}.

The fraction of correct answers is collected in Table~\ref{tab:benchmark_full}. The \ma architecture raises the total number of correct answers, most visibly on the fiducial-cut and off-shell questions. The SMEFT question is the sole exception. It is the most difficult of the five, and none of the configurations answers it reliably. The accumulated lessons improve the results further. The improvement is comparatively small, which we attribute to not having tailored the lessons to the specific questions. In practical workflows, where the accumulated lessons bear directly on the task at hand, we expect it to be much larger.

We also note that no configuration answers any question correctly in every run, and some runs fail for reasons unrelated to the designed silent failures. We believe that part of the remaining gap is set by the base model rather than by the \ma architecture. A stronger model, with better reasoning, coding, and instruction following, is expected to increase the fraction of correct answers and to answer the SMEFT question correctly. Finally, we also note that the results in Table~\ref{tab:benchmark_full} measure the correctness of an answer without any human review or intervention. In practice a physicist reviews the final setup and learns the actual \mg behavior.

\begin{table}[t]
  \centering
  \begin{small} \begin{tabular}{llccc}
    \toprule
    Test question & Bare & Cold & Warm \\
    \midrule
    Conjugate decays & 6/10 & 9/10 & 9/10 \\
    Decay nesting & 5/10 & 7/10 & 8/10 \\
    Fiducial cuts & 0/10 & 5/10 & 5/10 \\
    Off-shell decay & 0/10 & 2/10 & 5/10 \\
    SMEFT setup & 0/10 & 1/10 & 0/10 \\
    \bottomrule
  \end{tabular} \end{small}
  \caption{Fraction of correct answers per question. The questions and the three setups are defined in the text.}
  \label{tab:benchmark_full}
\end{table}

\clearpage
\section{Repeatable Benchmark}
\label{sec:benchmark}

For our repeatable validation benchmark we inject new physics into our \sfitter dataset. The task of the agent is to recover its Wilson coefficients. Running the same \sfitter analysis over statistically independent pseudo-datasets measures how consistently it does that. Our signal is a massive coloron~\cite{Bai:2018jsr} that couples strongly to quarks, including top quarks,
\begin{align}
    \mathcal{L} \;\supset\; g_c\, C^a_\mu\, \bar q\, \gamma^\mu T^a q 
    \qquad \mwith \qquad 
    g_c = g_s \tan\theta \eqc
    \label{eq:coloron_coupling}
\end{align}
where $C^a_\mu$ denotes the massive color-octet vector field, and $q$ the quark fields, summed implicitly over all six flavors and color. The coupling $g_c$ is set by the strong coupling $g_s$ and a free mixing angle $\tan\theta$. We choose the model parameters
\begin{align}
    M_c=3.75~~\tev \qqquad 
    \tan\theta=2.1 \qqquad 
    \Gamma_c =1.26~~\tev
\end{align}
This allows us to treat the coloron through SMEFT while also allowing the coloron to remain clearly detectable. To be consistent with our dataset we use the Run~2 energy and luminosity,
\begin{align}
13~~\tev 
\qqquad 
140.1~\text{fb}^{-1}
\end{align}
%

\begin{figure}[b!]
    \centering
    \includegraphics[width=0.70\linewidth]{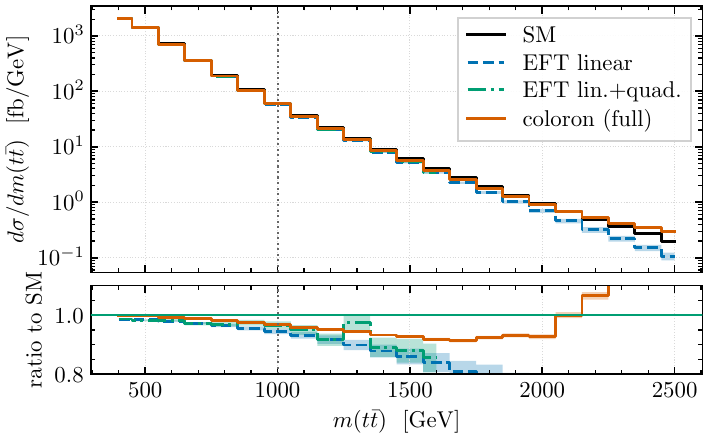}
    \caption{Three-way SMEFT-matching validation in $m(\ttbar)$, comparing the Standard Model, SM + dimension-six SMEFT at linear order, SM + SMEFT at linear and quadratic order, and the full coloron sample, each with per-curve MC statistical band. 
    }
    \label{fig:coloron_benchmark}
\end{figure}

To validate the \sfa results, we need the correctly matched Wilson coefficients. Integrating out the coloron at tree level generates six color-octet two-light-two-heavy four-quark operators all sharing the same Wilson coefficient
\begin{align}
    \frac{c_{Qq}^{(8,1)}}{\Lambda^2} 
    = \frac{c_{Qu}^{(8)}}{\Lambda^2} 
    = \frac{c_{Qd}^{(8)}}{\Lambda^2} 
    = \frac{c_{tq}^{(8)}}{\Lambda^2} 
    = \frac{c_{tu}^{(8)}}{\Lambda^2} 
    = \frac{c_{td}^{(8)}}{\Lambda^2} 
    \equiv \frac{c_8}{\Lambda^2} 
    = -\frac{g_c^2}{M_c^2} = - \frac{0.40}{~\tev^2}\;
    \eqc
    \label{eq:c8}
\end{align}
where the minus sign reflects the destructive interference of the coloron with the Standard Model $\ttbar$ amplitude. This matching could be done more precisely at higher-order perturbation theory, but this approximate result is sufficient to check the numerical \sfitter result. Because the coloron is a pure vector, it leaves the top charge asymmetry untouched, $A_C=0$, which provides us with another test.

A three-way comparison of the Standard Model, the Standard-Model-plus-SMEFT prediction, and the full coloron sample, shown in Figure~\ref{fig:coloron_benchmark}, confirms the picture at the level of the total rate.
The dimension-six prediction, including the quadratic term, tracks the coloron to about $1.3~\tev$ in $m(\ttbar)$.
Above that the full samples cannot be brought to overlay at any single value of $c_8$, since a dimension-six contact interaction cannot reproduce a Breit-Wigner resonance once the invariant mass approaches the pole.
This is the expected breakdown of the effective description and sets the validity window of the benchmark.

For full reproducibility we record every simulation choice used to build the benchmark in Table~\ref{tab:benchmark_sim}.
\begin{table}[t]
  \centering
  \begin{small} \begin{tabular}{@{}ll@{}}
    \catrule{software} \\ \addlinespace[2pt]
    \madgraph~\cite{Alwall:2014hca}        & $3.7.1$ \\ \addlinespace[3pt]
    \textsc{LHAPDF}~\cite{Buckley:2014ana} & $6.5.5$ \\ \addlinespace[3pt]
    Python                                 & $3.13.5$ \\ \addlinespace[3pt]
    Python packages                        & \textsc{numpy} $2.3.5$, \textsc{scipy} $1.18.0$, \textsc{matplotlib} $3.10.7$, \textsc{PyYAML} $6.0.3$ \\
    \addlinespace[5pt]
    \catrule{models} \\ \addlinespace[2pt]
    coloron                 & \texttt{ThetaColoron2025\_UFO}, \textsc{FeynRules}~\cite{Alloul:2013bka} version $1.4.6$ \\ \addlinespace[3pt]
    dimension-six operators & \texttt{dim6top\_LO\_UFO}~\cite{BessidskaiaBylund:2016jvp}, dated 2020-05-19, \texttt{gauge=[0,1]} \\ \addlinespace[3pt]
    two additional scalars  & $m_{\Theta_N}=5~\tev$, $m_{\phi_I}=10~\tev$, couplings vetoed at generation level \\
    \addlinespace[5pt]
    \catrule{physics inputs} \\ \addlinespace[2pt]
    PDF set         & leading-order \textsc{NNPDF23\_lo\_as\_0130\_qed}~\cite{Ball:2013hta} \\
                    & (\texttt{nn23lo1}, \texttt{lhaid=230000}) \\ \addlinespace[3pt]
    top pole mass   & $m_t=173.0~\gev$ (Yukawa mass $172.0~\gev$) \\ \addlinespace[3pt]
    strong coupling & $\alpha_s(m_Z)=0.1184$, from the PDF set \\
    \addlinespace[5pt]
    \catrule{generation} \\ \addlinespace[2pt]
    scale choice        & \madgraph default dynamical, event by event \\
                        & (\texttt{dynamical\_scale\_choice=-1}, \texttt{scalefact=1.0}) \\ \addlinespace[3pt]
    jet flavors         & \texttt{maxjetflavor=4} \\ \addlinespace[3pt]
    Breit-Wigner window & \texttt{bwcutoff=15.0} \\ \addlinespace[3pt]
    simulation level cuts   & parton level only, \textsc{Pythia8}/\textsc{FastJet}/\textsc{HepMC} unused \\
  \end{tabular} \end{small}
  \caption{Simulation setup used to generate the benchmark samples. The dimension-six operators are imported alongside the coloron model for the same $\ttbar$ final state.}
  \label{tab:benchmark_sim}
\end{table}
The two interference samples and the two auxiliary SMEFT samples are each generated at $\mathcal{O}(10^6)$ events.

We turn the coloron sample into blind datasets in HEPData-like format, designed to look like genuine LHC differential $\ttbar$ data with no reference to the coloron or the SMEFT hypothesis.
Each dataset consists of two measurements that we shipped as self-contained, HEPData-like submissions, with the following respective observables:
\begin{itemize}
    \item the absolute differential cross section in $m(\ttbar)$, in eight bins from $345$ to $3~\tev$, reported with its statistical covariance, a $1.5\%$ fully correlated luminosity uncertainty, and a Standard Model prediction. 
    \item the charge asymmetry $A_C(m(\ttbar))$, in four bins from $345$ to $2,5~\tev$, reported with its statistical covariance and a Standard Model prediction. 
\end{itemize}
For the first dataset, we derive both measurements directly from the simulated data.
We generate an additional five datasets by Poisson bootstrapping the simulated events and using this to build additional measurements. 
Further, we independently draw luminosity shifts for each of these five datasets. 
This luminosity is shared between both measurements in a given dataset. 

This gives us a total of six datasets which are the input to our repeatability checks.

\subsubsection*{Repeated global analyses}
\label{sec:repeated}

\begin{figure}[t]
    \centering
    \includegraphics[width=\linewidth]{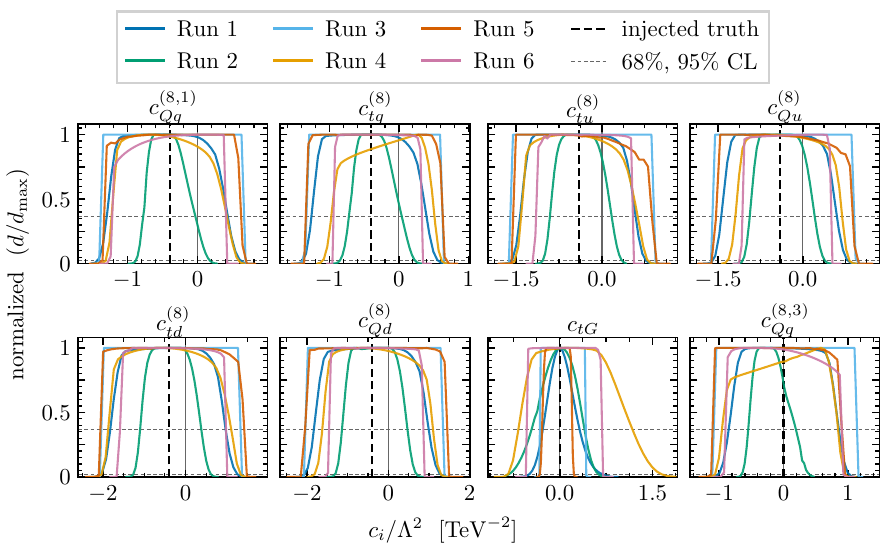}
    \caption{Joint profile likelihoods of the six repeated global analyses, one curve per run, for
    each of the eight Wilson coefficients. Every curve is the per-operator joint profile, with the
    remaining seven coefficients profiled over.}
    \label{fig:v103_profiles}
\end{figure}

We use the six datasets defined in the previous section to run six global analyses using \sfa.
We instruct the agent to perform an SFitter analysis on the two observables in each set in isolation, skipping the first part of the data extraction by directly providing the agent the data in the HEPData format. 
Since we deliberately designed \sfa as an interactive co-working tool, every agent run is supervised and driven by a physicist, starting with the unified prompt:
\begin{userblock}[User]
There is a new directory (run\_00) in your output path with two preliminary HEPData entries of an upcoming ttbar analysis. We’ve been asked to run SFitter on both of them together in isolation, meaning we only use that data in the directory, not the other top data cards that might exist. This is a genuine physics test, so if the results differ from the Standard Model, report it honestly, but apply standard caution for truly problematic bins. Both SM reference and data are at LO, so treat them like you normally would proper Full Simulation in relation to experimental data. 
Use both available observables. 
We are currently on our CPU cluster, so do all the preparations here and tell me when to switch to the GPU cluster for the final analysis.
\end{userblock}
The details of each agent run diverge following this. 
We supervised each run the way a reasonable physicist would. 
When asked for the operator set that should be included in the global analysis, we choose the six color-octet operators, along with $\ope_{tG}$ and $\ope_{Qq}$, as our $t\bar{t}$ measurement is sensitive to these operators.
Likewise, we guide the choice of uncertainties for each agent. 
Finally, we want to report the results in the form of a collection of joint-analysis profiled and marginalized single-operator limits. 
Therefore, we specifically instruct the agents to produce their outputs in this shape in cases where the default output differs from it. 

\begin{figure}[t]
    \centering
    \includegraphics[width=\linewidth]{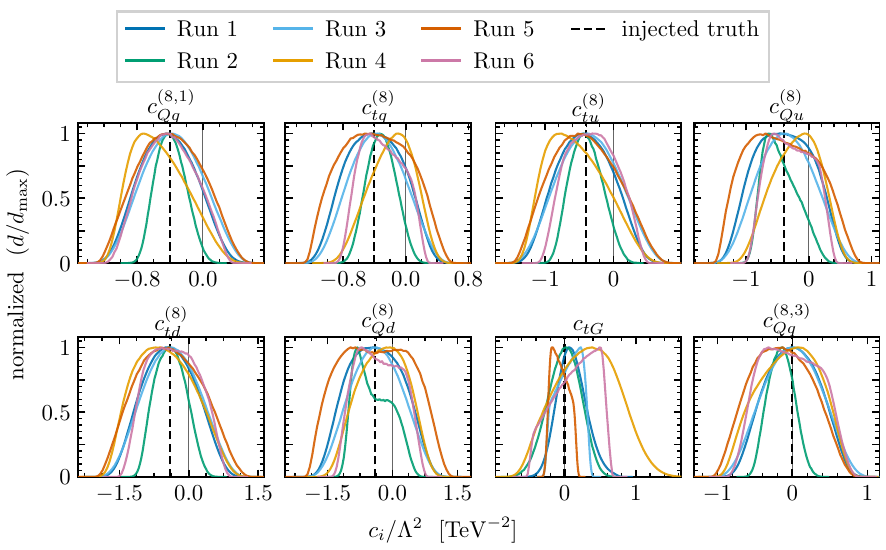}
    \caption{Marginalized likelihoods of the six runs and the eight Wilson coefficients,
    each normalized to its own peak, with the injected truth shown as the dashed line.
    The y-axis is the over-peaked $L^2$ posterior of the tool, not a frequentist interval.}
    \label{fig:v103_marginals}
\end{figure}

The profile results of each global analysis are shown in Figure~\ref{fig:v103_profiles}. Marginals are shown in Figure~\ref{fig:v103_marginals}.
For each figure, we show all 8 relevant Wilson coefficients and compare them to the truth value, indicated by the dashed line.  
In the profile collection, we see significant flat regions in each likelihood.  
This is a genuine result of the degeneracy between the six color-octet operators.
\sfa are able to break this effect in the marginal case, resulting in significantly tighter constraints centered around the true value.
Across both figures, we see good agreement between individual agent runs, accounting for statistical fluctuations in the bootstrapped data sets and the finite-sized simulation each agent ran.

An arguable outlier is Run 2, which managed to go beyond the original task and understood that we were attempting to perform a synthetic test. 
The agent in Run 2 therefore concluded that the correct scale uncertainty was zero, as the SM reference and data were generated with the same settings, except for the injected signal. 
When prompted to produce a version of the global analysis with a more realistic scale uncertainty band, the previous finding about the nature of the data still caused it to lean toward a narrower uncertainty band than other runs. 
In this specific case, the agent made no mistake; a global analysis with no scale uncertainty does indeed lead to the best results in this scenario. 
We therefore do not view this outlier as a failure. 

\subsubsection*{Reproducible documentation}

Clear documentation that allows for the direct reproduction of older results is a key feature of any physics analysis. 
It is also an ideal use case for agentic assistance tools.
Therefore, we specifically defined a strict set of documentation instructions and requirements in \sfa. 

The documentation consists of 5 subdirectories associated with each internal workflow step and an overarching Markdown file. 
The Markdown file contains a summary of the global analysis and describes the content of each subfolder.
The contents of each subfolder are partially predefined and partially chosen by the agent. 
Each directory is required to have a \texttt{code/} folder that contains every piece of code the agent used. Likewise required is a Markdown file that meticulously lists every decision step the agent or user took and documents the contents of \texttt{code/}. 
Other subdirectories, such as \texttt{validation/} in \texttt{datacard/} are  managed by the agent. 
If the agent decides a subdirectory with, e.g., validation plots is useful for documenting its work, it creates one such self-managed folder. 
An example structure of the produced documentation is:

\dirtree{%
.1 documentation/.
.2 datacard/.
.3 code/.
.3 validation/.
.3 datacard.json.
.3 datacard.md.
.2 extraction/.
.3 code/.
.3 processed/.
.3 raw/.
.3 extraction\_report.json.
.3 extraction.md.
.2 fit/.
.3 code/.
.3 results/.
.3 fit.md.
.2 parametrization/.
.3 code/.
.3 processed/.
.3 parametrization.md.
.2 sm\_template/.
.3 code/.
.3 runcards/.
.3 sm\_template.md.
.2 README.md.
}

When prompted to re-generate the result of a previous global analysis from this documentation structure, the agent can reproduce the previous results exactly.

\bibliographystyle{tepml}
\bibliography{tilman,refs, dataset}

\end{document}